\documentclass[twocolumn,superscriptaddress,amsmath,amssymb,aps,pra]{revtex4-2}
\usepackage{svg}
\usepackage{graphicx}
\usepackage{amsmath}
\usepackage{amssymb}
\usepackage{physics}
\usepackage{braket}
\usepackage{hyperref}
\usepackage{xcolor}
\usepackage{xspace}
\usepackage{tikz}
\usetikzlibrary{arrows.meta, positioning}
\definecolor{magicpurple}{HTML}{B793FF}
\usepackage{comment}
\usepackage{algorithm}
\usepackage{algpseudocode}
\usepackage[normalem]{ulem}
\usepackage{enumitem}
\usepackage[section]{placeins} 
\usepackage{subcaption}

\newcommand{\signal}{\ensuremath{S_{\mathrm{PA}}(\theta)}\xspace}
\newcommand{\sigmaSQsc}{0.02} 
\newcommand{\sigmaZZsc}{0.127}
\newcommand{\sigmaSQwh}{0.02} 
\newcommand{\sigmaZZwh}{0.117}
\newcommand{\sigmaZZdwh}{0.125}

\newcommand{\colorsquare}[1]{\textcolor[HTML]{#1}{\rule{0.8em}{0.8em}}}
\begin{document}

\title{Observation of gravity-like signatures in holographic codes on a quantum computer}

\author{Debopriyo Biswas}
\thanks{These authors contributed equally to this work.}
\affiliation{Duke Quantum Center, Duke University, Durham, NC 27701 USA}
\author{Gong Cheng}
\thanks{These authors contributed equally to this work.}
\affiliation{Virginia Tech Center for Quantum Information Science and Engineering, Department of Physics, Virginia Tech, Blacksburg, VA 24061, USA}
\author{Krishnanand Karthikeyan}
\thanks{These authors contributed equally to this work.}
\affiliation{Virginia Tech Center for Quantum Information Science and Engineering, Department of Physics, Virginia Tech, Blacksburg, VA 24061, USA}
\author{Diana Muñoz-Valencia}
\affiliation{National Quantum Laboratory (QLab), Joint Quantum Institute and Department of Physics, University of Maryland, College Park, MD 20742 USA}

\author{Vincent P. Su}
\affiliation{BlueQubit Inc., San Francisco, CA 94105}

\author{Hrant Gharibyan}
\affiliation{BlueQubit Inc., San Francisco, CA 94105}

\author{Daiwei Zhu}
\affiliation{IonQ, Inc., College Park, MD  20740  USA}

\author{Grant Salton}
\affiliation{IonQ, Inc., College Park, MD  20740  USA}

\author{Evgeny Epifanovsky}
\affiliation{IonQ, Inc., College Park, MD  20740  USA}

\author{Martin Roetteler}
\affiliation{IonQ, Inc., College Park, MD  20740  USA}

\author{Christopher Monroe}
\affiliation{Duke Quantum Center, Duke University, Durham, NC 27701 USA}
\affiliation{IonQ, Inc., College Park, MD  20740  USA}

\author{John Preskill}
\affiliation{Institute for Quantum Information and Matter, California Institute of Technology, Pasadena, CA 91125 USA}

\author{Norbert M. Linke}
\affiliation{National Quantum Laboratory (QLab), Joint Quantum Institute and Department of Physics, University of Maryland, College Park, MD 20742 USA}
\affiliation{Duke Quantum Center, Duke University, Durham, NC 27701 USA}

\author{ChunJun Cao}
\affiliation{Virginia Tech Center for Quantum Information Science and Engineering, Department of Physics, Virginia Tech, Blacksburg, VA 24061, USA}

\author{Crystal Noel}
\affiliation{Duke Quantum Center, Duke University, Durham, NC 27701 USA}

\begingroup

\begin{abstract}
The unification of quantum mechanics and general relativity remains one of the major open problems of theoretical physics. The Anti-de Sitter/Conformal Field Theory (AdS/CFT) correspondence provides a valuable theoretical framework for this effort via a holographic duality between a theory of quantum gravity in asymptotically AdS spacetime and a conformal quantum field theory on the lower-dimensional boundary. Here, we implement a toy model of this duality called the HaPPY code~\cite{Pastawski:2015qua}, a quantum error-correcting code in the form of a tensor network with hyperbolic entanglement patterns, on a trapped-ion quantum computer. We present the first experimental confirmation of the Faulkner--Lewkowycz--Maldacena formula in this model---a key test of the holographic correspondence~\cite{Ryu_2006,flm}. We then enrich it with non-stabilizerness, or magic, and observe entropic precursors expected of emergent gravity. 
Finally, we present and measure a code construction whose entropic behavior is reminiscent of a highly quantum wormhole~\cite{Maldacena:2013xja}. 
Our experiments illustrate how quantum computers can serve as testbeds for modeling the emergence of spacetime.
\end{abstract}

\endgroup


\maketitle

\section*{Introduction}

A central question in quantum gravity is whether spacetime is fundamental or instead emerges from more elementary quantum degrees of freedom. Although a complete theory of quantum gravity remains elusive, several leading approaches ---including string theory and loop quantum gravity---increasingly suggest that spacetime is an emergent phenomenon rather than a fundamental ingredient of nature~\cite{VanRaamsdonk:2010pw,Pastawski:2015qua,Swingle:2009bg,Faulkner_2014,Jacobson_2016,Cao_2017, Maldacena_1999,Witten:1998qj,Jacobson_1995,Oriti:2006se, cotler2019entanglement}. The most concrete realization of this idea is the AdS/CFT correspondence~\cite{Maldacena_1999,Witten:1998qj}, which relates a bulk quantum theory with gravity to an equivalent boundary quantum theory without gravity. Within this duality, spacetime geometry of the bulk theory is encoded in the quantum entanglement of the boundary theory. This connection is quantified by the Ryu--Takayanagi (RT) and Faulkner--Lewkowycz--Maldacena (FLM) formulae, which relate the entanglement entropy of a boundary region to geometric quantities in the bulk gravitational description~\cite{Ryu_2006,Ryu:2006bv,flm}.

Quantum error-correcting codes (QECCs) provide a simple setting in which these ideas can be explored~\cite{Almheiri_2015,Harlow:2016vwg,Pastawski:2015qua, Hayden_2016, Dong:2016eik, Cao_2018,Akers_2019, Cao_2021,Qi_2022, Akers:2021fut,Dong_2019}. In these models, bulk reconstruction is analogous to quantum error correction~\cite{Almheiri_2015,Dong:2016eik, cotler2019entanglement, chen2020entanglement}, allowing holographic duality to be studied in finite-dimensional quantum systems. A paradigmatic example is the HaPPY (Harlow--Pastawski--Preskill--Yoshida) code~\cite{Pastawski:2015qua}, a tensor-network construction whose entanglement structure reproduces the RT formula and key features of the holographic dictionary. Yet despite their success, existing holographic codes largely describe fixed background geometries and do not capture how quantum matter modifies the emergent geometry --- a feature expected in gravity.

Recent advances in programmable quantum computers have created new opportunities to investigate simplified models of quantum gravity experimentally. Works such as Refs.~\cite{Sachdev_1993, Kitaev:2015talk3,brown2023quantum, nezami2023quantum,Schuster_2022,Sahay:2024vfw,Vikram:2026wdg} propose the implementation of holographically inspired models on quantum platforms. Recent experiments \cite{Landsman2019, jafferis2022traversable, shapoval2023towards} have explored wormhole-inspired teleportation protocols, information scrambling, and many-body chaos. An NMR test of the RT formula (without the quantum correction of FLM) on 6-qubit perfect tensors was also demonstrated by Ref.~\cite{Li_2019}. However, these demonstrations do not directly investigate how quantum matter influences emergent geometry.

\begin{figure*}[h!tbp]
\centering
\includegraphics[width=\textwidth]{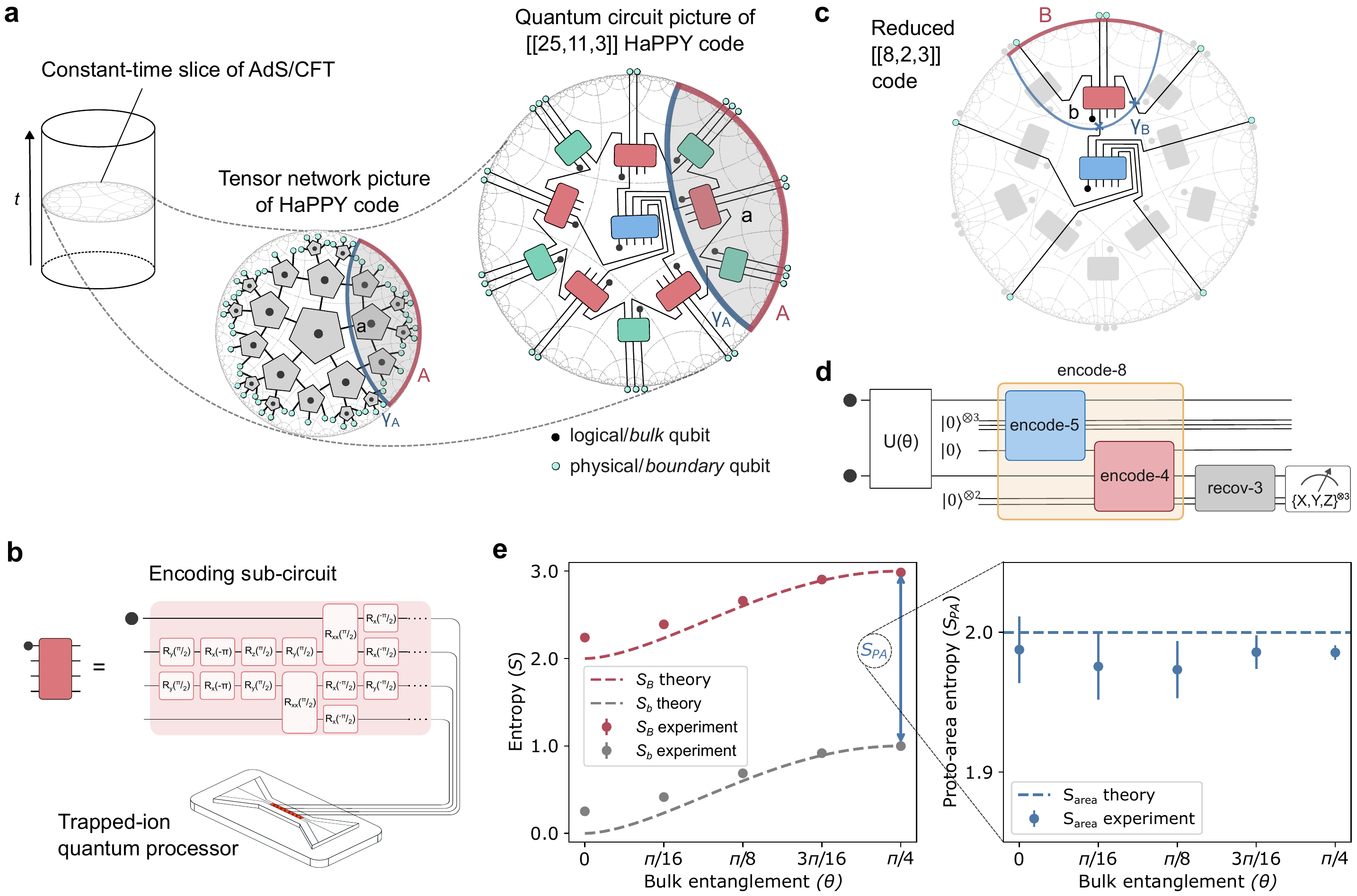}
\caption{\textbf{Experimental realization of a holographic quantum error correcting code and FLM formula test.}
\textbf{a}:  A hyperbolic time slice of AdS is tessellated by perfect tensors and compiled into a quantum circuit, with the encoding blocks running radially outward to the boundary. Blue, red, and green rectangles are the three distinct encoding sub-circuits (Appendix~\ref{app:codeconstruction}); black and cyan dots are the logical (\textit{bulk}) and physical (\textit{boundary}) qubits respectively, and the remaining inputs are initialized to $\ket{0}$. $A$ is an example boundary subregion and $\gamma_A$ is the minimal-area surface homologous to it.
\textbf{b}: Decomposition of an encoding block into $\{R_x, R_y, R_z, R_{XX}\}$ gates for the trapped-ion processor (Appendix~\ref{sec:transpilation}).
\textbf{c}: We choose $A$ to be a five-qubit subregion. The code is then reduced to an $[[8,2,3]]$ code where $A$ becomes a connected 3-qubit boundary region $B$, with minimal-area bulk surface $\gamma_B$ (blue).
\textbf{d}: Circuit for (c): two logical qubits are entangled by $U(\theta)$ [Eq.~\eqref{eq:U_theta}], encoded, recovered on $B$, and read out by quantum state tomography (Appendix~\ref{app:reduction}).
\textbf{e}: Test of the FLM relation. Left: measured boundary entropy $S_B$ and recovered bulk entropy $S_b$ versus the entangling angle $\theta$, compared with ideal-theory curves. Right: the proto-area entropy $S_{\mathrm{PA}}$ [Eq.~\eqref{eq:PAentropy}].  2100 shots are taken per Pauli basis; error bars are from bootstrapping.
} 
\label{fig:rt}
\end{figure*}

Here we realize holographic quantum error-correcting codes on a programmable trapped-ion quantum computer and use them to experimentally probe the relation between bulk matter and emergent geometry. Trapped-ion processors are particularly well suited to holographic encoding circuits because of their native all-to-all qubit connectivity. We study the two-layer $[[25,11,3]]$ HaPPY code (Fig.~\ref{fig:rt}a) and develop explicit encoding and local recovery circuits tailored to the trapped-ion gate set. Exploiting the isometric structure of the tensor network, we construct and implement reduced circuits on the IonQ \textit{Forte} quantum computer with access to up to 36 qubits~\cite{forte_benchmarking}, which faithfully reproduce the entropic observables of the full code while substantially reducing the experimental resources required.

To probe the interplay between bulk matter and geometry, we reconstruct both the reduced boundary state $\rho_A$ and the optimally recoverable bulk state $\rho_a$ for a connected boundary subregion $A$ (Fig.~\ref{fig:rt}a) using quantum state tomography (QST). We characterize the emergent geometry through the proto-area entropy~\cite{cao2026},
\begin{equation}\label{eq:PAentropy}
S_{\mathrm{PA}}(A)=S(\rho_A)-S(\rho_a),
\end{equation}
which measures the difference between the boundary entanglement entropy and the entropy of the recoverable bulk degrees of freedom.

\begin{figure*}[tb]
\centering
\includegraphics[width=\textwidth]{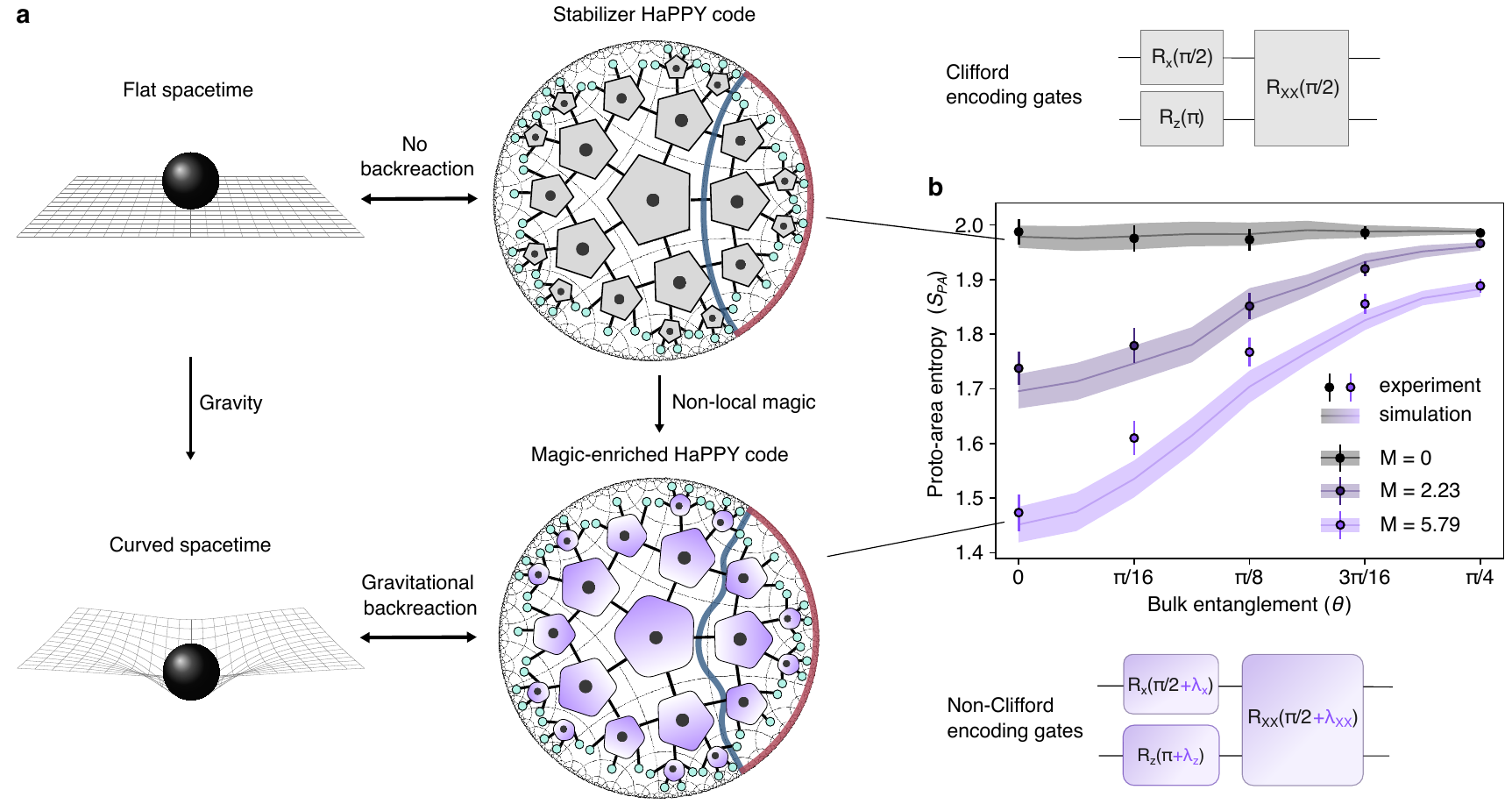}
\caption{
\textbf{Magic-enriched holographic codes and gravitational precursor.}
\textbf{a}: Conceptual and gate-level schematic of (top) an exact stabilizer HaPPY code and (bottom) a magic-enriched HaPPY code. The former is constructed from Clifford encoding gates and realizes a fixed tensor-network geometry, analogous to a rigid background with no gravitational backreaction. Introducing controlled coherent over-rotations $\lambda_i$ to the encoding gates adds magic (non-stabilizerness) to the latter code, producing a state-dependent proto-area entropy, analogous to gravitational backreactions shifting the underlying geometry. 
\textbf{b}: Proto-area entropy $S_{\mathrm{PA}}$ measured as a function of the bulk entanglement angle $\theta$ for three values of added magic. $M$ denotes the total magic of the perturbed encoding circuit, quantified using stabilizer R\'enyi entropy. Experimental data for $M=0$ includes 2100 shots per Pauli basis, while $M>0$ uses 1500 shots. Entropies are estimated by MLE and error bars are $\pm1\sigma$ estimated by bootstrapping. 
}
\label{fig:sc}
\end{figure*}
We first apply this protocol to the original stabilizer HaPPY code, where the RT and FLM relations are expected to hold exactly. The measured proto-area entropy agrees with the FLM prediction within experimental uncertainty, providing the first experimental verification of the FLM formula in a holographic quantum error-correcting code. 

We then extend the construction beyond the stabilizer setting by introducing non-Clifford (``magic") resources into the encoding circuit, allowing us to investigate how bulk entanglement modifies the emergent geometry.
Unlike the stabilizer code, its proto-area (PA) entropy depends on the bulk entanglement. In a setting that mimics a single AdS geometry, we find that $S_{\mathrm{PA}}$ increases with bulk entanglement, a behavior expected in holographic theories with gravity. In a two-copy code that entangles the bulk (logical) qubits of the two sides---analogous to entangling two AdS spacetimes---we observe a decrease in $S_{\mathrm{PA}}$. This may be analogous to the change of holographic geometries in the presence of a wormhole, but the observed effect in the toy model should not be taken as a literal probe of semi-classical geometry.

\section*{Faulkner--Lewkowycz--Maldacena formula test} \label{sec:RT}

The RT formula and its quantum refinement, the FLM relation \cite{Ryu_2006,flm}, are key entries in the holographic dictionary. 
The entanglement wedge (EW) of boundary subregion $A$ is the region enclosed by a so-called minimal (or extremal) surface $\gamma_A$ in the bulk and $A$ (shaded region in Fig.~\ref{fig:rt}a). 
The FLM relation relates the area of $\gamma_A$ to the entropy $S_A$ and the entropy $S_a$ of the bulk degrees of freedom in the EW of $A$:
\begin{equation} \label{eq:flm}
    S_A = \frac{\mathrm{Area}(\gamma_A)}{4G_N} + S_a,
\end{equation}
where $G_N$ is the gravitational constant. 

For clarity, we refer to the leading order entropy term $S_\text{area}=\frac{\mathrm{Area}(\gamma_A)}{4G_N}$ as the area term, which contains geometric information of the bulk. 
The HaPPY code realizes this relation concretely---both the geometry and the encoding map are realized as a tensor-network, which we translate into a quantum circuit (Fig.~\ref{fig:rt}a). The minimal cut in the HaPPY tensor network, equal to $S_{\rm PA}$, then plays the role of the area term \cite{Pastawski:2015qua,Harlow:2016vwg,cao2026}. Since the code can correct for the erasure of $A^c$ (the complement of $A$) exactly, the bulk state $\rho_a$ also coincides with the pre-encoded logical state when restricted to the logical qubits in the EW of $A$. 

In our first experiment, we test the FLM relation for a connected $5$-qubit boundary subregion $A$ of the full $[[25,11,3]]$ HaPPY code. By entangling the bulk qubit at the center with the one closest to it in the EW of $A$, $S_a$ will increase because the recoverable bulk qubit in the EW of $A$ will become more mixed. If the FLM formula holds, then $S_A$ must also increase in a way such that $S_{\rm area}$ is always equal to 3 --- the number of edge cuts. 

Leveraging the isometric properties of the tensor network, we can reduce the 25-qubit system to an entropically equivalent description tailored to the subregion $A$. This reduced $[[8,2,3]]$ code with an effective $3$-qubit boundary subregion $B$ (Fig.~\ref{fig:rt}c) fully preserves the desired bulk entropy dependence in the original EW of $A$ even though the expected ``area contribution'' $S_{\rm area}= S_{\rm PA}(B)$ is reduced from 3 to 2. We show in Appendix~\ref{app:reduction} that this reduction does not impact the key signals we aim to verify in this paper. 

To verify the FLM relation experimentally, we prepare bulk states with different amounts of entanglement, encode them, then measure the corresponding bulk and boundary entropies. The procedure is as follows: 
\begin{enumerate}[leftmargin=10pt]
\item Prepare two qubits in an entangled Bell-like state using a unitary $U(\theta)$ such that
\begin{equation}
\label{eq:U_theta}
U(\theta)\ket{00}=\cos{\theta}\ket{00}+\sin{\theta}\ket{11},
\end{equation} 
\item Encode these two qubits by applying the encoding circuit as shown in Fig.~\ref{fig:rt}d.
\item Apply a recovery circuit on qubits in subregion $B$.  
\item Perform quantum state tomography after recovery to extract the boundary entropy $S_B$ and recovered bulk entropy $S_b$ to calculate $S_{\mathrm{PA}}(B)$ as in Eq.~\eqref{eq:PAentropy}.
\end{enumerate}

We extract both the bulk and boundary entropies after applying the recovery circuit in Fig.~\ref{fig:rt}d because entropy is invariant under local unitary rotations in $B$. This is also crucial for canceling out systematic errors in a noisy circuit (Appendix~\ref{app:noisysystematics}).

The experimental results from IonQ hardware and the theory predictions are shown in Fig.~\ref{fig:rt}e.
We ran $3^3=27$ copies of each circuit, each with $23~ZZ$ gates, to measure the 3 boundary qubits in each of $\{X, Y, Z\}$ Pauli bases. Maximum Likelihood Estimation (MLE) is used to reconstruct both the boundary and bulk entropies. 
As $\theta$ is varied, the bulk entropy $S_b$ is predicted to grow from zero at $\theta=0$ to one at $\theta=\pi/4$ when the recoverable bulk state on $B$ is maximally mixed. The boundary entropy should follow the same trend but exactly offset by $2$. The experimental measurement confirms this FLM prediction where the ``area'' contribution is fixed when the bulk entanglement is varied.

\section*{Magic-enriched single-copy code} \label{sec:SC}

Contrary to the FLM test above, the area contribution should depend on the bulk state in a truly gravitational theory. Changes in the bulk entropy are tied to changes in the stress energy \cite{Faulkner_2014,Jacobson_2016,Cao_2018, Lashkari:2013koa, Blanco:2013joa,Czech:2016tqr}, which, through Einstein's equations, backreact on the geometry and shift the minimal surface areas \cite{Dong_2019,cao2026}. A similar dependence follows from the quantum extremal surface (QES) prescription~\cite{Engelhardt_2015} expected in a theory of gravity, in which the surface $\gamma_A$ is chosen by extremizing the generalized entropy. As such, the area term itself becomes state-dependent. We therefore expect a model of emergent gravity to show a similar response, in which changes in the bulk matter state backreact on the geometry (Fig.~\ref{fig:sc}a).

The absence of such dependence in the HaPPY code can be traced to its stabilizer nature --- gravitational backreaction is expected to require non-stabilizer resources, in particular non-local magic \cite{Cao:2023mzo, Cao:2024nrx, cao2026}. This requires us to consider a magic-enriched version of the HaPPY code as a minimal setting in which emergent gravitational behavior may arise.

To inject magic into the HaPPY code, we introduce predetermined over- and under-rotations into the gates of the encoding circuit in the $\{R_x, R_y, R_z, R_{XX}\}$ basis. 
Each single- and two-qubit gate angle is offset from its Clifford value by $\lambda_i$ radians. These offsets control the amount of injected magic $M$, with larger values of $\lambda$ corresponding to higher magic (details in Appendix~\ref{app:sec:magic_injection}). As an example, the three values of $M$ shown in Fig.~\ref{fig:sc}b correspond to over-rotation angles $\lambda_{XX}=\{0,0.1,0.2\}~\mathrm{rad}$ for the $R_{XX}$ gates. The full set of offsets is listed in Table~\ref{tab:lambdas_823}. 
With these perturbations, the encoding circuit becomes magic-enriched and no longer defines an exact stabilizer code. We quantify the resulting magic $M$ using the stabilizer R\'enyi entropy of the code's Choi state (Appendix~\ref{app:NLmagic}).

We consider the case in which one of the encoding sub-circuits remains an isometry after magic injection. Under this condition, the 25-qubit magic-enriched HaPPY code again reduces to a magic-enriched \( [[8,2,3]] \) code (Appendix~\ref{app:reduction}). 

The experimental procedure for probing $S_{\text{PA}}$ in the magic-enriched codes follows the same four-step structure as the FLM formula test, with two key differences. First, the exact HaPPY encoding circuit is replaced by a magic-enriched version. 
Second, since the QECC now only corrects erasures approximately, Ref.~\cite{cao2026} proposes that an optimal  recovery circuit acting on $A$ that minimizes the trace distance between the recovered state and the input logical state for each instance of the magic-enriched encoding needs to be applied to recover the bulk state. 
However, such a recovery requires a deeper and thus noisier circuit that is less cost-effective for practical implementation. 

To reduce the two-qubit gate count, we therefore restrict the recovery circuit to have the same circuit architecture as the Clifford recovery, and optimize only within this fixed ansatz. Details of the optimization algorithm are provided in Appendix~\ref{app:recovery}. From the measured boundary and recovered bulk entropies, $S_B$ and $S_b$, we compute $S_{\mathrm{PA}}$ defined in Eq.~\eqref{eq:PAentropy} and use it as an analog of area in the emergent geometry. Note that the constraint-induced suboptimality influences the strength of the state-dependence. 

Since the constrained recovery could in principle produce a spurious state dependence, in Appendix~\ref{app:optimal} we numerically compute $S_{\mathrm{PA}}$ using the unconstrained optimal recovery. The state dependence persists, albeit with reduced magnitude, confirming that the signal is not an artifact of the recovery ansatz.

\begin{figure}[tb]
\centering
\includegraphics[width=\columnwidth]{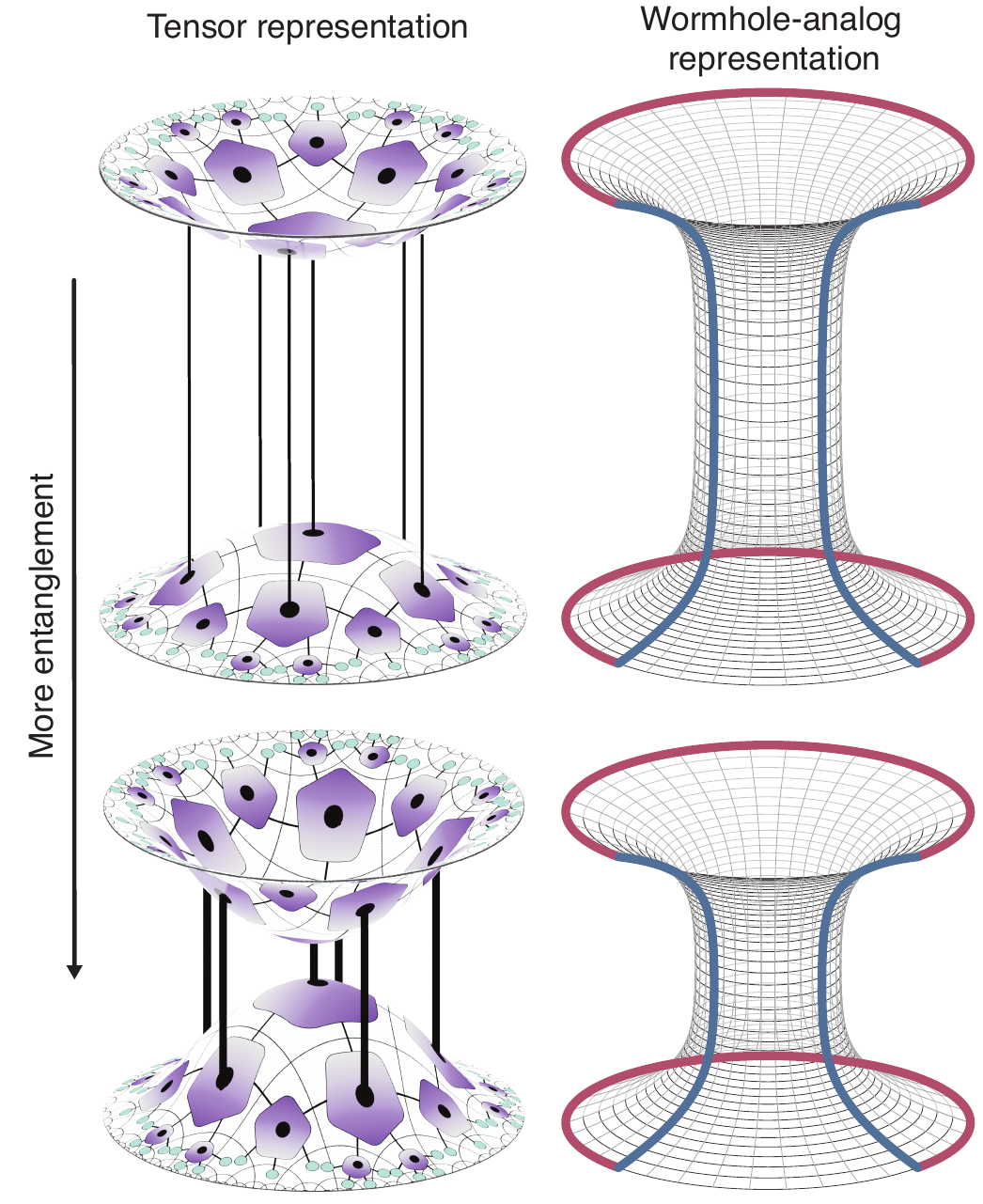}
\caption{
\textbf{Entanglement-induced connectivity.}
Two magic-enriched HaPPY-code tensor networks are coupled through entanglement between corresponding logical degrees of freedom (top left). In the wormhole analogy (top right), boundary-anchored geodesics, shown in blue, pass through the neck connecting the two sides. (Bottom) increasing the inter-code entanglement decreases the effective wormhole length, thereby shortening this candidate geodesic and reducing the geometric contribution to the entropy in sufficiently short wormholes in holography. Red curves denote the boundary subregions. }
\label{fig:wh_schematic}
\end{figure}
The experimental results for three different values of injected magic are shown in Fig.~\ref{fig:sc}b. The $M=0$ data-points correspond to the proto-area entropy from the FLM test. For the injected-magic ($M>0$) experiments we use the same data collection and entropy reconstruction protocols as the previous experiment. We observe that $S_{\mathrm{PA}}$ increases with the bulk entanglement, realizing an important state-dependent relation expected in a theory with emergent gravity~\cite{cao2026}.  
The sensitivity of the proto-area entropy to $\theta$ also increases with both the total magic $M$, as seen from the increasing slope, and the non-local magic of the codewords (Appendix~\ref{app:NLmagic}), consistent with expectations from holography~\cite{Cao:2024nrx}. 

We compare the experimental data to quantum circuit simulation under noisy conditions with stochastically fluctuating gate angles which represent the dominant error source in this system \cite{forte_benchmarking}. Single- and two-qubit gate angles are modeled as fluctuating from their intended value by a Gaussian distribution with standard deviations of \{$\sigma_{1q}$, $\sigma_{2q}$\} = \{\sigmaSQsc, \sigmaZZsc\}~radians, matching independent measurement of gate fidelities.
Simulation details for each experiment are in Appendix \ref{sec:noisynumerics}.
 
\section*{Double-sided entangling experiment} \label{sec:WH}

\begin{figure*}[t]
\centering
\includegraphics[width=1\textwidth]{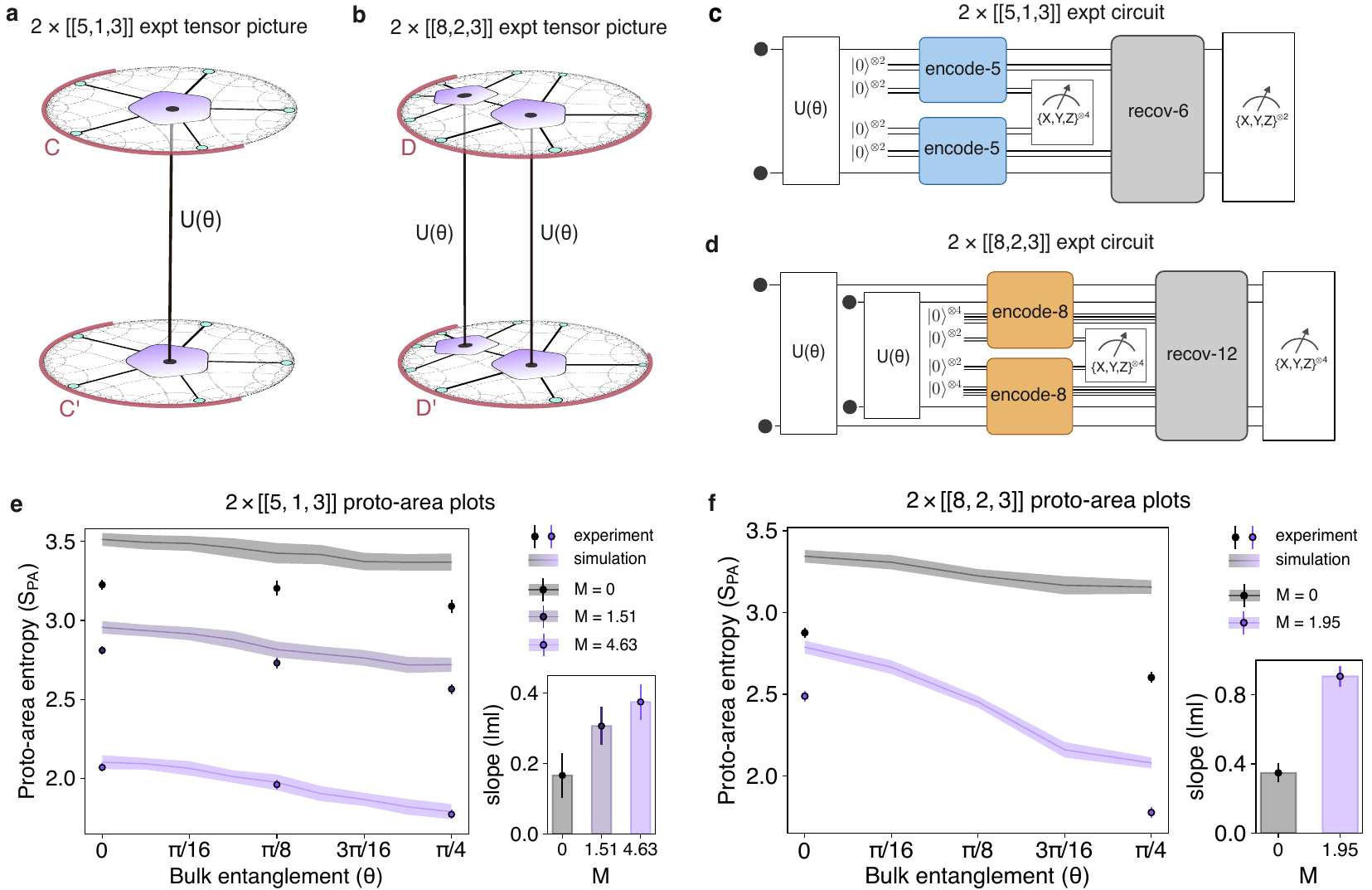}
\caption{
\textbf{Double-sided experiments from entangled holographic codes.}
\textbf{a,b}: Tensor-network diagrams for the two-code wormhole analogue. The $10$-qubit experiment entangles two $[[5,1,3]]$ codes using their logical qubits, while the $16$-qubit experiment extends to entangling two $[[8,2,3]]$ codes using two pairs of logical qubits.
\textbf{c,d}: Corresponding experimental circuits used to extract boundary and bulk entropies.
\textbf{e,f}: Measured proto-area entropy $S_{\mathrm{PA}}$ as a function of the bulk entanglement angle $\theta$ for the $10$-qubit and $16$-qubit experiments. Experimental data are compared with numerical simulations. The $10$-qubit data were acquired with $1250$--$1500$ shots per Pauli basis, and the $16$-qubit data were acquired with $1000$ shots per Pauli basis. Entropies are estimated by MLE and error bars are $\pm1\sigma$ estimated by bootstrapping. The accompanying bar plots show the magnitude $|m|$ of the slope obtained from weighted linear fits to the experimental $S_{\mathrm{PA}}(\theta)$ data for each value of total injected magic $M$.
}
\label{fig:wh_data}
\end{figure*}

If spacetime geometry truly emerges from entanglement, this notion should be testable through the Einstein--Rosen\,=\,Einstein--Podolsky--Rosen (ER=EPR) proposal \cite{Maldacena:2013xja,VanRaamsdonk:2010pw}. In this proposal, entanglement is not merely a property of states on a fixed background but is a key ingredient in building spacetime connectivity. Increasing the entanglement between two otherwise disconnected systems can make their dual spacetime more connected, for instance through wormhole-like structures \cite{VanRaamsdonk:2010pw,Maldacena:2013xja,Qi:2013caa,Pastawski:2015qua,Cao_2017}. Establishing this theoretically is challenging in general \cite{Haehl:2019fjz}, but a direct experimental probe can potentially provide evidence and guide future theory.

We test this idea with two disjoint copies of the (magic-enriched) HaPPY code, approximately dual to two disconnected AdS geometries (Fig.~\ref{fig:wh_schematic}). If entanglement acts as the ``glue'' of geometry, as Refs.~\cite{VanRaamsdonk:2010pw,Maldacena:2013xja} propose, then more strongly entangled codes should appear more geometrically connected. Rather than entangling boundary degrees of freedom into a conventional thermofield-double state \cite{Maldacena:2001kr}---dual to a global wormhole background in holographic CFTs---we instead entangle the \textit{bulk} degrees of freedom. This provides a more direct probe of how bulk matter entanglement can backreact on the emergent geometry, and is motivated by broader proposals that extend beyond the standard AdS/CFT setting \cite{VanRaamsdonk:2010pw,Maldacena:2013xja,Cao_2017,Hayden_2016}. Moreover, while well-motivated from the ER=EPR proposal, this bulk entangled construction does not have a well understood gravitational dual; our goal is to shed light on such a dual through our experiment. 

The quantity of interest is the proto-area entropy $S_{\mathrm{PA}}$ of a combined boundary region, formed by selecting one boundary subregion from each of the two code blocks. $S_{\mathrm{PA}}$ again plays the role of the area of the minimal surface in the entanglement geometry. 
Without injected magic, we expect that $S_{\mathrm{PA}}$ should be state-independent and equal to the sum of the two codes' separate ``area'' contributions. In other words, the entropy of the union is the sum of the two minimal cuts from two disjoint systems. 
However, we predict qualitatively distinct behavior once magic is introduced: $S_{\mathrm{PA}}$ develops a nontrivial dependence on the inter-code entanglement, reflecting not only an additional bulk-entropy contribution but also a modification of the effective geometric connectivity between the two sides. Although the precise geometry of this deformation is unclear, we can intuit the change of $S_{\mathrm{PA}}$ using known results on the holographic wormhole~\cite{Maldacena:2013xja,VanRaamsdonk:2010pw,Cao_2017,Morrison:2012iz} (Appendix~\ref{app:Entanglement and wormhole length}). In contrast to the single-code case, where $S_{\mathrm{PA}}$ increases with bulk entropy, the wormhole intuition suggests that it will decrease as the entanglement between the two codes is increased. 

We test these hypotheses on two double-sided constructions with different sizes. 
The smaller realization is a $10$-qubit model built from two copies of $[[5,1,3]]$ codes, shown schematically in Fig.~\ref{fig:wh_data}a, with the corresponding circuit in Fig.~\ref{fig:wh_data}c. The larger realization starts from two $[[25,11,3]]$ HaPPY codes and, using the same code reduction method, reduces to an effective $16$-qubit two-sided configuration where each side is described by a reduced $[[8,2,3]]$ code. This is shown schematically in Fig.~\ref{fig:wh_data}b, with the corresponding circuit in Fig.~\ref{fig:wh_data}d. 
Connected subregions are drawn from the two codes: two $3$-qubit regions (one per side) for the $10$-qubit construction forming a combined boundary region $C\cup C'$, and two $6$-qubit regions for the effective $16$-qubit construction forming a combined boundary region $D\cup D'$. 
In both cases, the two HaPPY codes are coupled by entangling the logical qubits across the two sides using the same unitary $U(\theta)$ as before in Eq.~\eqref{eq:U_theta}. The logical inputs are then encoded into their respective boundary blocks. The 10-qubit construction encodes one Bell pair, whereas the 16-qubit construction encodes two; we therefore expect the probed signal to be more pronounced in the latter case. 
To reduce experimental costs of QST, we obtain the boundary entropy of the smaller complementary region and use the fact that the global state is pure in the limit of zero noise to compute the boundary entropies of $C\cup C'$ or $D\cup D'$. This important difference from the single-copy experiment introduces $\theta$-dependent systematics even in the zero-magic circuit (Appendix~\ref{app:noisysystematics}), therefore a nonzero slope here cannot be attributed purely to magic injection. Nevertheless, an increasing $\theta$ dependence with injected magic  still provides an independent check on magic-induced
state dependence. Further implementation details are provided in Appendix~\ref{app:twosided}. 

The circuits are magic-enriched by shifting the gate angles away from Clifford rotation angles, as in the single-copy experiment. In the $2\times[[5,1,3]]$ experiment, the coherent offsets are chosen in the range $\lambda_i \in [0, 0.27]$ radians, with larger offsets corresponding to higher total magic (specific $\lambda$ choices explained in Appendix~\ref{app:sec:wh10_magic_injection}). For the $2\times[[8,2,3]]$ experiment, we optimize the $\lambda_i$ gate by gate within the range of $[-0.10, 0.10]$ radians to enhance the \signal signal. The optimized $\lambda_i$ gives total magic $M=1.95$. In the $M=0$ realization, all offsets are set to $\lambda_i = 0$ (details in Appendix~\ref{app:sec:wh16_magic_injection}).

Similar to the single-copy experiment, a fully unconstrained optimal recovery circuit is costly on hardware. For the two-sided $[[8,2,3]]$ code, identifying the fully unconstrained optimal recovery is computationally expensive even in theory, let alone implementing it on quantum hardware. Therefore, we adopt an experimentally feasible recovery ansatz sharing the same circuit architecture as the Clifford recovery, and optimize the recovery within this restricted ansatz.

The results of the experiments on IonQ hardware are shown in Fig.~\ref{fig:wh_data} e-f. In the $2\times[[5,1,3]]$ experiment, we ran $3^4+3^2=90$ copies of each circuit to measure the bulk and boundary entropies respectively. Each circuit has up to $31$ $ZZ$ gates. In the $2\times[[8,2,3]]$ experiment it is $3^4+3^4=162$ circuit copies, each with up to $56$ $ZZ$ gates.
Density matrices are reconstructed using MLE, and the error bars in Fig.~\ref{fig:wh_data}e,f denote $\pm1\sigma$ uncertainties estimated by bootstrapping. 
Our simulations use the same noise model as the single-copy experiment, with standard deviations of these stochastic gate angle fluctuations given by \{$\sigma_{1q}$, $\sigma_{2q}$\} = \{\sigmaSQwh, \sigmaZZwh\}~radians.
However, simulation with stochastic error alone does not match the experimental data very well, and we study the effects of additional coherent errors as a possible explanation in Appendix \ref{sec:noisynumerics}.

In both realizations, the simulations reproduce the experimentally observed decrease of $S_{\mathrm{PA}}$ with increasing $\theta$ when magic is injected. 
We compare the results of differing amounts of magic by reporting the slopes $|m|$ of weighted linear fits to the data. 
 
The magnitude of slopes increases with injected magic, indicating that the reduction in proto-area entropy becomes more pronounced with added magic and more bulk entanglement.


Since the model has no known gravitational dual, there is no first-principles prediction for the proto-area entropy in this setting; our experiment therefore serves as an exploratory probe and motivates further theoretical investigation. As a consistency check, we numerically compute $S_{\mathrm{PA}}$ for the two-sided [[5,1,3]] setup using the unconstrained optimal recovery and find that the signal persists, albeit with reduced magnitude (Appendix~\ref{app:optimal}). The observed effect is therefore not an artifact of the restricted recovery ansatz.

\section*{Discussion and Outlook}

The results presented here lay the foundation for future explorations of quantum gravity in the lab. 
We show that by creating experimentally feasible constructions, we can observe gravity-like signatures that build intuition for new models of quantum gravity that may not yet have a satisfying theoretical description.
For example, the study of the quantum ``proto-wormhole'' we present here may be helpful for future theoretical developments involving the ER=EPR conjecture. 

For scalable future studies using entropic probes, we also highlight the importance of developing more efficient tomography schemes such as near-stabilizer or tensor-network-informed tomography \cite{Cramer:2010lwr,QSTbdd_extent}, where the cost need not grow exponentially. 
Equipped with such resource-efficient tomography techniques, a scaled-up experiment with $3\times$ more qubits and higher fidelity gates would also inform the conditions needed to emerge semi-classical geometry \cite{Cao_2020}. 

Identifying the optimal recovery protocol is another potential bottleneck for scalability — indeed, in the present small-scale experiment, suboptimal recovery substantially amplifies the measured state dependence (Appendix~\ref{app:optimal}). However, Ref.~\cite{Witten:2026gze} argues that the effect of recovery suboptimality is strongly suppressed in the large-system limit where semi-classical geometry is expected to emerge, so this obstacle should become less severe precisely in the regime of greatest interest.
Studies of dynamical systems where geometry can be learned from correlation functions without tomography may also reduce the resource requirements as the system scales to larger qubit numbers \cite{Engelhardt_2016,Sahay:2024vfw}.

More broadly, these experiments point toward a future in which programmable quantum hardware serves as a laboratory for quantum gravity---probing how spacetime emerges from entanglement in regimes no classical computer can reach.

\FloatBarrier

\begin{acknowledgments}
We thank Aidan Chatwin-Davies, Ananth Kaushik, and Brian Swingle for helpful discussions. This work is supported by a collaboration between the US Department of Energy and other Agencies. This material is based upon work supported by the U.S. Department of Energy, Office of Science, National Quantum Information Science Research Centers, Quantum Systems Accelerator (award no. DE-SCL0000121); and by the Department of Energy through QuantISED awards DE-SC0019380 and DE-SC0018407. This work was funded by the National Science Foundation, Software-Tailored Architectures for Quantum Co-Design (STAQ) award (PHY-2325080) and QLCI: Center for Robust Quantum Simulation (OMA-2120757).
Access to IonQ Forte was provided by the National Quantum Laboratory (QLab) at the University of Maryland.

The authors declare the following competing interests: D.Z., G.S., E.E., M.R. and C.M. are employees of IonQ, Inc. and have personal financial interests in the company. J.P. is a shareholder and part-time employee of Oratomic, Inc., which is developing fault-tolerant quantum computers. C.C. serves as a paid consultant to Microsoft.
\end{acknowledgments}

\section*{Author contributions}
C.C. and C.N. conceptualized the project.
C.C., G.C., K.K., V.S., H.G., J.P. developed the theoretical framework.
C.C., G.C., K.K., and V.S. developed the circuit design.
C.C., G.C., and K.K. developed measurement protocol and established the theoretical interpretation of gravity-like signatures.
D.B., G.C., K.K., C.C., C.N., V.S., D.Z., E.E. developed the methodology. 
D.B. led the experimental implementation and data analysis.
D.B., G.C., K.K. performed numerical simulation relevant to the experiment. 
D.B. and D.M.-V. performed experiments on the IonQ Forte system and developed software toolkits for experiments.
G.C., K.K., D.B., D.Z., J.P., N.M.L., C.C., C.N., G.S., C.M. contributed to interpretation of results.
D.B., K.K., D.M-V., G.S., and G.C. produced all figures.
N.M.L., C.N., C.M., M.R. supervised experiments, provided resources, and acquired funding. 
C.C., J.P. supervised the theoretical development and acquired funding. 
All authors contributed to writing and reviewing the manuscript.

\bibliographystyle{unsrt}
\bibliography{references}

\clearpage
\appendix
\onecolumngrid
\section{HaPPY Code and circuit construction}\label{app:codeconstruction}
\subsection{Encoding circuit construction}
\begin{figure*}[t]
    \centering
    \includegraphics[width=\linewidth]{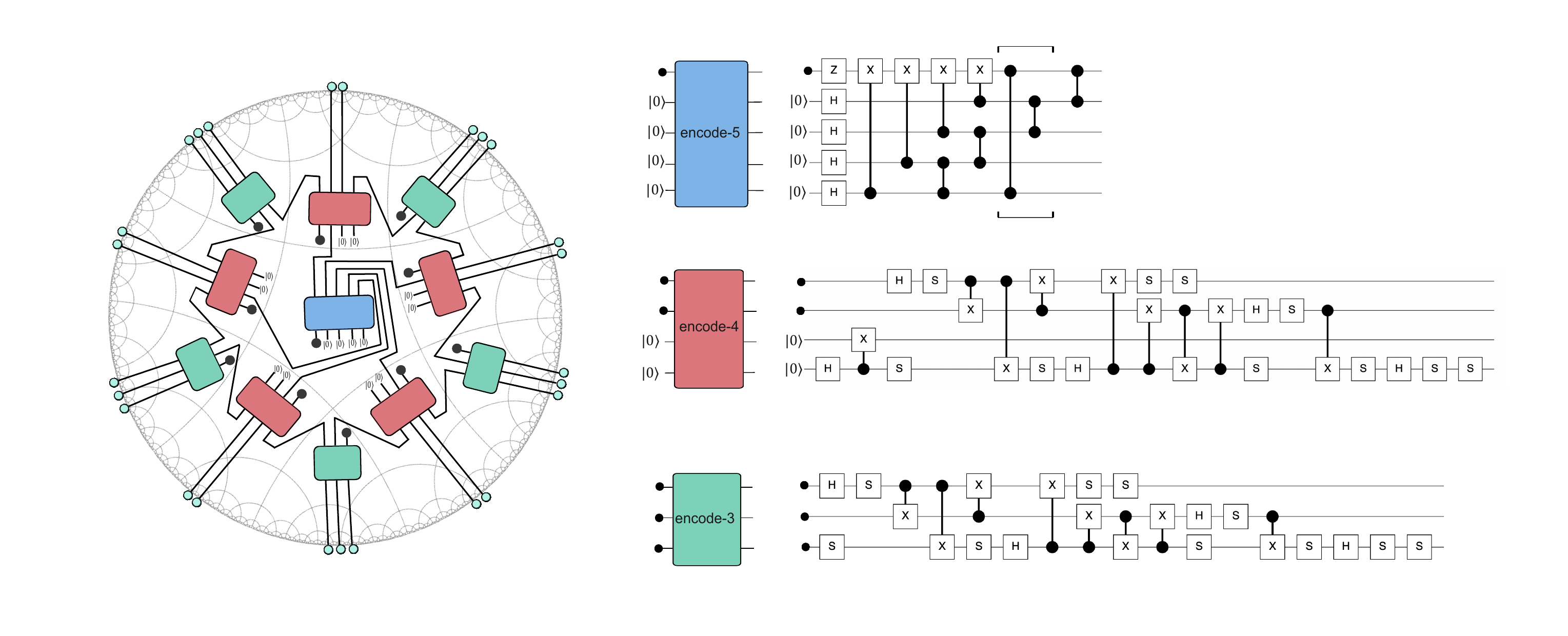}
    \caption{\textbf{Two-layer HaPPY encoding network and local circuit primitives.}
Left, tensor-network geometry of the two-layer HaPPY code used to construct the 25-qubit encoding circuit from logical bulk qubits to physical boundary qubits. The central blue tensor is interpreted as a $[[5,1,3]]$ encoding block, the surrounding red tensors as $[[4,2,2]]$ encoding blocks, and the outer green tensors as $[[3,3,1]]$ unitary blocks. Teal dots on the boundary denote the physical output qubits. Right, circuit realizations of the three local building blocks used in the construction: the $[[5,1,3]]$ encoding unitary, the $[[4,2,2]]$ encoding unitary, and the $[[3,3,1]]$ unitary block. Composing these local maps according to the tensor-network connectivity gives the full 25-qubit bulk-to-boundary encoding circuit.}
    \label{fig:happy25_circuit}
\end{figure*}
The 25-qubit circuit is constructed from the tensor network of the two-layer HaPPY code~\cite{Pastawski:2015qua}, consisting of a central tensor and two surrounding layers. The encoding circuit is obtained by code concatenation: each layer of perfect tensors is interpreted as an encoding map associated with a local seed code.

\begin{figure}[H]
    \centering
\includegraphics[width=0.75\linewidth]{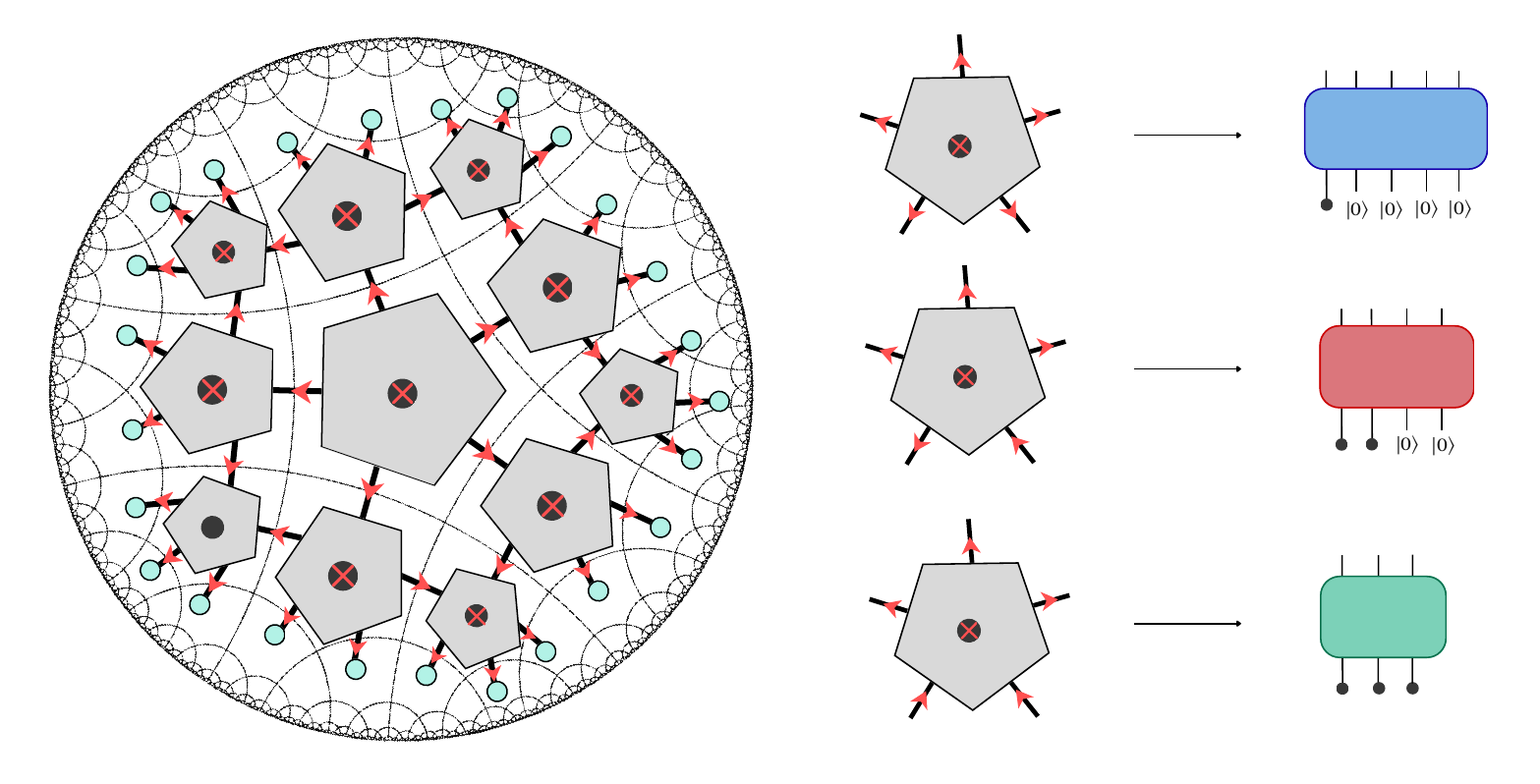}
    \caption{\textbf{Circuit representations of the local tensor isometries.}
A 6-leg perfect tensor can be interpreted as different encoding maps depending on the choice of input and output legs. The three choices used in constructing the two-layer HaPPY encoding circuit are shown: the $[[5,1,3]]$ seed code, the $[[4,2,2]]$ seed code, and the $[[3,3,1]]$ unitary block. Black dots denote logical input legs, while legs initialized in $\ket{0}$ denote ancilla inputs.}
    \label{fig:tensor_to_codes}
\end{figure}

Because the elementary tensor (defined by the 5-qubit code) is perfect, its legs can be partitioned in different ways to define different isometries, and hence different quantum codes.  The starting point is the $[[5,1,3]]$ code, whose encoding unitary maps one logical qubit (together with four ancilla qubits initialized in $\ket{0}$) to five physical qubits. Converting one physical tensor leg to a logical leg produces the second tensor (a $[[4,2,2]]$ error-detecting code) corresponding to an encoding unitary mapping two logical qubits (with two ancilla qubits) to four physical qubits. Moving another physical leg to logical results in a three-to-three unitary that we represent as a $[[3,3,1]]$ code; a unitary map from three logical qubits to three physical qubits. All three cases are shown in Fig.~\ref{fig:tensor_to_codes}.

To construct the 25 qubit HaPPY code, we proceed from the center outward. The central tensor is first viewed as a $[[5,1,3]]$ code, encoding the central logical qubit into five qubits. These five qubits are then concatenated with five $[[4,2,2]]$ codes in the next layer: each output qubit from the central $[[5,1,3]]$ code is paired with one new logical qubit in the second layer and encoded into four output qubits. Two of those output qubits correspond to genuine physical qubits on the boundary, while the other two output qubits are then contracted with the outer layer of the code. Specifically, we use five tensors that correspond to encoding some qubits into $[[3,3,1]]$ codes. For each, a pair of adjacent $[[4,2,2]]$ blocks contributes one contracted qubit each; together with one new logical qubit in the outer layer, these three qubits are mapped with the encoding unitary of the $[[3,3,1]]$ code to three physical qubits. Thus, the boundary comprises three physical qubits from each of the five $[[3,3,1]]$ tensors, and two physical qubits from each of the five $[[4,2,2]]$ tensors, for a total of $15+10=25$ physical qubits. We refer to the encoding circuits for the $[[5,1,3]]$, $[[4,2,2]]$, and $[[3,3,1]]$ codes as \textit{encode-5}, \textit{encode-4}, and \textit{encode-3}, respectively. The contracted tensor network defines the encoding circuit from the bulk, logical qubits to the boundary, physical qubits. 

\subsection{Magic injection}\label{app:sec:magic_injection}
To inject magic into the encoding circuit, we add a persistent over- or under-rotation offset to each gate, as shown in Eq.~\eqref{eq:gates1} and Eq.~\eqref{eq:gates2}. While the encoding circuit we construct is a Clifford circuit, these coherent over- and under-rotations result in encoding gates with the rotation angles that are shifted away from their Clifford values (\(\pi/2\) for the single-qubit rotations and \(\pi/4\) for the two-qubit rotations), resulting in a non-Clifford encoding circuit. We parametrize these coherent offsets by angles \(\lambda_i\). For single-qubit gates, we use rotations about the $x$ and $y$ axes, 
\begin{equation}
R_x(\phi) := e^{-i\phi X/2}, 
\qquad
R_y(\phi) := e^{-i\phi Y/2}.
\end{equation}

The ideal Clifford gates appearing in the encoding circuit are then skewed by adding small rotation offsets $\lambda_i$ to the single- and two-qubit gate angles. For example,
\begin{align}\label{eq:gates1}
X &= R_x\!\left({\pi}\right) \;\to\; R_x\!\left({\pi}+\lambda_x\right),\\
Y &= R_y\!\left({\pi}\right) \;\to\; R_y\!\left({\pi}+\lambda_y\right),
\end{align}
and
\begin{equation}
H = R_x\!\left({\pi}\right) R_y\!\left(\frac{\pi}{2}\right)
\;\to\;
R_x\!\left({\pi}+\lambda_x\right)
R_y\!\left(\frac{\pi}{2}+\lambda_y\right).
\end{equation}
Similarly, for the two-qubit entangling gates we use the parametrization
\begin{equation}
R_{XX}(\phi) = e^{-i\phi X\otimes X/2}.
\end{equation}
At $\phi=\pi/2$, this gate is Clifford and generates maximal entanglement. We inject magic by shifting this rotation angle away from its Clifford value,
\begin{equation}\label{eq:gates2}
 R_{XX}\!\left(\frac{\pi}{2}\right)
\;\to\;
R_{XX}\!\left(\frac{\pi}{2}+\lambda_{XX}\right).
\end{equation}

In our implementation, the values of $\lambda_i$ are picked and held fixed throughout each experiment. Their values are summarized in Tables~\ref{tab:lambdas_823} and~\ref{tab:lambdas_10wh}.

\subsection{Circuit transpilation to native gates} \label{sec:transpilation}

\begin{figure}[t]
    \centering

    \newlength{\flowblockwidth}
    \setlength{\flowblockwidth}{0.68\textwidth}
    \newcommand{\compilationfigscale}{0.75}

    \scalebox{\compilationfigscale}{%
    \begin{tikzpicture}[
        font=\sffamily\small,
        node distance=0.95cm,
        layer/.style={
            draw,
            rounded corners=4pt,
            line width=0.4pt,
            minimum width=\flowblockwidth,
            minimum height=1.25cm,
            inner sep=0pt
        },
        flowarrow/.style={
            -{Latex[length=3mm]},
            thick
        },
        textblock/.style={
            align=center,
            text width=\dimexpr\flowblockwidth-2.3cm\relax,
            inner sep=0pt
        }
    ]

    \newcommand{\gateicon}[1]{%
        \begin{tikzpicture}[baseline=-0.6ex, every node/.style={outer sep=0pt}]
            \node[
                draw,
                line width=0.7pt,
                minimum height=0.40cm,
                inner sep=1.5pt
            ] (g) {$#1$};
            \draw[line width=0.7pt] ([xshift=-0.22cm]g.west) -- (g.west);
            \draw[line width=0.7pt] (g.east) -- ([xshift=0.22cm]g.east);
        \end{tikzpicture}%
    }

    \node[layer] (Cirq) {};
    \node[layer, below=of Cirq] (Qiskit) {};
    \node[layer, below=of Qiskit] (magic) {};
    \node[layer, below=of magic] (ionq) {};

    \node[anchor=west] at ([xshift=0.16cm]Cirq.west) {%
        \gateicon{H}
    };

    \node[anchor=west] at ([xshift=0.16cm]Qiskit.west) {%
        \gateicon{R_x\!\left(\frac{\pi}{2}\right)}
    };

    \node[anchor=west] at ([xshift=0.16cm]magic.west) {%
        \gateicon{R_x\!\left(\frac{\pi}{2}\!+\!{\color{magicpurple}\lambda_x}\right)}
    };

    \node[anchor=west] at ([xshift=0.16cm]ionq.west) {%
        \gateicon{G_{\pi}({\color{magicpurple}\phi})}
    };

    \node[textblock] at ([xshift=0.65cm]Cirq.center) {%
        Basis gates: $H$, $Z$, $S$, $\mathrm{CX}$\\[2pt]
        \emph{Cirq representation}
    };

    \node[textblock] at ([xshift=0.65cm]Qiskit.center) {%
        Basis gates: $R_x(\theta_i)$, $R_y(\theta_i)$, $R_z(\theta_i)$, $R_{XX}(\theta_j)$\\[2pt]
        \emph{Qiskit representation}
    };

    \node[textblock] at ([xshift=1.0cm]magic.center) {%
        Basis gates: 
        $R_x(\theta\!+\!{\color{magicpurple}\lambda_x})$, 
        $R_y(\theta\!+\!{\color{magicpurple}\lambda_y})$, 
        $R_z(\theta\!+\!{\color{magicpurple}\lambda_z})$, 
        $R_{XX}(\theta\!+\!{\color{magicpurple}\lambda_{XX}})$\\[2pt]
        \emph{Qiskit representation, Non-Clifford gates}
    };

    \node[textblock] at ([xshift=0.65cm]ionq.center) {%
        Native gates: 
        $GPi({\color{magicpurple}\varphi})$, 
        $GPi2({\color{magicpurple}\varphi})$, 
        $ZZ({\color{magicpurple}\vartheta})$\\[2pt]
        \emph{Qiskit in IonQ Forte native gates}
    };

    \draw[flowarrow] (Cirq.south) -- (Qiskit.north);
    \draw[flowarrow] (Qiskit.south) -- (magic.north);
    \draw[flowarrow] (magic.south) -- (ionq.north);

    \end{tikzpicture}%
    }

    \caption{Circuit compilation flow from the original Cirq representation to IonQ native gates.}
    \label{fig:circuit-compilation-flow}
\end{figure}
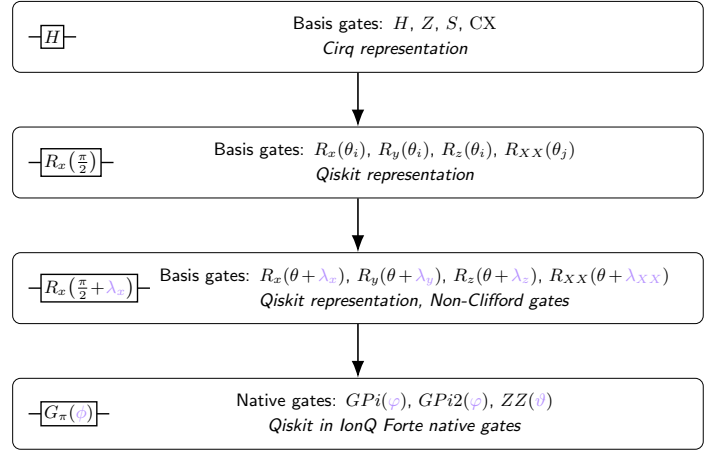

As illustrated in Fig.~\ref{fig:circuit-compilation-flow}, the magic-free HaPPY-code circuits are first written in the Clifford gate set $\{H,Z,S,CX\}$, which is a natural representation of the underlying stabilizer code. We then convert these circuits to the gate set $\{R_x,R_y,R_z,R_{XX}\}$, whose rotation angles can be varied continuously. This representation allows us to introduce magic in a controlled way by shifting selected gate angles away from their Clifford values. For example, an intended rotation $R_x(\theta)$ is replaced by $R_x(\theta+\lambda_i)$, where $\lambda_i$ is the corresponding angle offset in radians.

The resulting circuits are then transpiled to the native gate set of the IonQ Forte-1 quantum computer using a custom Python package developed for this work~\cite{ionq-gate-translator}. On Forte-1, the native single-qubit operations are the phase-parameterized gates GPi$(\varphi)$ and GPi2$(\varphi)$, corresponding to $\pi$- and $\pi/2$-pulses about an axis set by $\varphi$, respectively. The native two-qubit gate is a $ZZ(\vartheta)$ gate, where $\vartheta$ defines the entangling phase between the two qubits. The compiler uses standard decompositions of non-$R_z$ operations into native GPi, GPi2, and ZZ gates, while $R_z$ rotations are implemented virtually by tracking each qubit's rotating-frame phase and absorbing the accumulated phase into the parameters of subsequent native gates.

For Clifford circuits, our custom transpiler produced native-gate circuits with at least $60\%$ fewer single-qubit gates than the version of the Qiskit IonQ Provider used for comparison~\cite{qiskit_ionq}. For non-Clifford circuits, the gate-count reduction was smaller. Consequently, the magic-enriched encoding circuits can contain more single-qubit gates than the corresponding zero-magic circuits. Based on the measured single-qubit gate fidelities and noisy numerical simulations, we do not expect this additional single-qubit gate count to measurably affect the \signal signal.

\subsection{Recovery optimization}\label{app:recovery}
Given an encoding circuit $V$ and a boundary subregion $A$, our goal is to construct a recovery channel that acts on the reduced state $\rho_A$ and reconstructs the bulk state $\rho_a$ contained in the entanglement wedge (EW) of $A$.

We construct the recovery channel by applying a unitary $R$ to the qubits in $A$, followed by tracing out a subset $A_2\subset A$, so that the remaining qubits have the same Hilbert-space dimension as the target logical state $\rho_a$. The sequence of encoding, restriction to $A$, and recovery defines a combined quantum channel $\mathcal{N}_R$. For an input bulk state $\rho$, this channel acts as
\begin{equation}
    \mathcal{N}_R(\rho):=
    \Tr_{A_2}\left(R\,\Tr_{A^c}\left(V\rho V^{\dagger}\right)R^{\dagger}\right),
\end{equation}
where $A^c$ represents the complementary region of $A$. For the original HaPPY code, a local Clifford recovery circuit can be found using the greedy algorithm~\cite{Pastawski:2015qua}, which results in a recovery that exactly reconstructs the bulk state in the entanglement wedge;
\begin{equation}
    \mathcal{N}_R(\rho)=\rho_a .
\end{equation}

For the magic-enriched code, an exact local recovery generally no longer exists. We therefore optimize the recovery circuit variationally, by minimizing the distance between the recovered state and the target bulk state. Specifically, we use gradient descent to optimize the recovery unitary $R$ with respect to the cost function
\begin{equation}\label{eq:cost}
    f(R):=\mathbb{E}_U\left\lVert \mathcal{N}_R(U\rho(\theta)U^{\dagger})-U\rho_a(\theta)U^{\dagger}\right\rVert_1,
\end{equation}
where $\lVert \cdot\rVert_1$ denotes the trace norm. The average is taken over local unitaries $U:=\otimes_{i=1}^{n_L} u_i$ acting locally on the $n_L$ bulk qubits, which change the local basis but preserve the bulk entanglement. 

 For the original HaPPY code, the cost function can be minimized to zero, and the resulting recovery unitary is equivalent to the exact Clifford recovery. For the magic-enriched code, by contrast, the recovery must be determined by numerically minimizing Eq.~\eqref{eq:cost}. We can obtain the unconstrained optimal recovery for the single-copy $[[8,2,3]]$ setup and the two-sided $[[5,1,3]]$ setup. However, the resulting recovery unitaries require substantial experimental resources and are difficult to implement on current quantum hardware. Moreover, for the two-sided $[[8,2,3]]$ setup, even finding the unconstrained optimum numerically is computationally demanding. We therefore use an experimentally feasible ansatz with the same architecture as the Clifford recovery and optimize only within this fixed ansatz.

\section{Experimental details and results}

\subsection{Reduced $[[8,2,3]]$ HaPPY Code}\label{app:reduction}
\begin{figure*}[t]
    \centering
    \includegraphics[width=\linewidth]{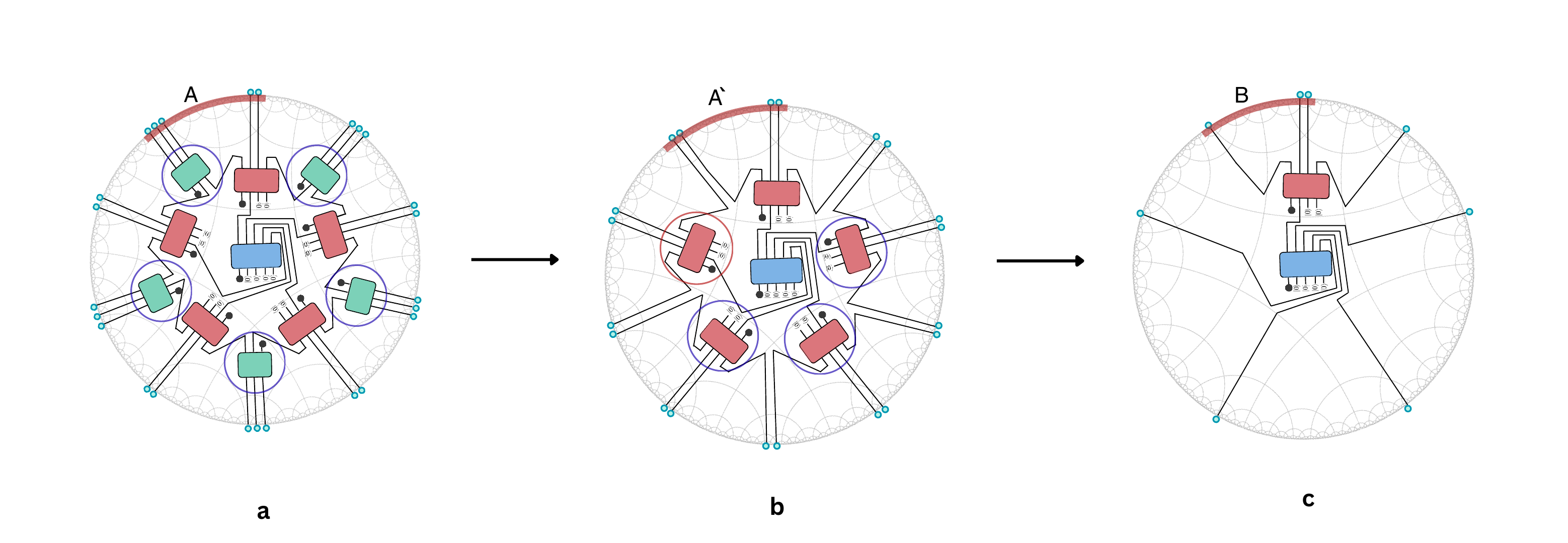}
\caption{\textbf{Isometric reduction of the boundary-entropy measurement.}
(a) Full two-layer \( [[25,11,3]] \) HaPPY network with the chosen five-qubit boundary region \(A\). Outer green tensors whose output legs lie entirely on one side of the bipartition act only as local unitaries and can be removed without changing the relevant entropy.
(b) The resulting intermediate network, where \(A\) is mapped to an effective boundary region \(A'\). Red tensors that act entirely within the complementary region can be removed, while the remaining red block contributes only a fixed Bell-pair entropy and can be absorbed into the reduction.
(c) The final entropy-equivalent reduced network, represented as an effective \( [[8,2,3]] \) circuit with boundary region \(B\). This reduced description preserves the entropy relevant to the protocol, with the ideal stabilizer area contribution reduced to \(2\) bits.}
    \label{fig:reduction}
\end{figure*}
\subsubsection{Encoding}
Taking advantage of the isometric structure of the HaPPY encoding circuit, the required entropy measurements can be performed using a reduced circuit that is equivalent to the full 25-qubit HaPPY circuit for the quantities of interest.

The reduction is illustrated in Fig.~\ref{fig:reduction}. We start from the full two-layer HaPPY network and choose five boundary qubits as the subregion $A$. Three of these qubits are output legs of the same green unitary. Because this unitary acts entirely within $A$, it does not change the eigenvalues of the reduced density matrix on $A$. We may therefore remove this block for the purpose of the entropy measurement. The same argument applies to the other green unitaries in the outer layer whose output legs lie entirely on one side of the bipartition.

After removing these outer green unitaries, the network reduces to an intermediate configuration consisting of the inner red unitaries and the central blue unitary. Similarly, three of the red unitaries (those circled in blue in Fig.~\ref{fig:reduction}) act entirely within the complementary region $A^c$, and therefore can also be removed without affecting $S(A)$. To further simplify the circuit, we consider a special case of magic enrichment for the red-circled red block such that it is an isometry from the single qubit in $A$ to the three qubits in $A^c$. Equivalently, by applying a local unitary on the qubits in $A^c$ that will be traced out, one can distill a Bell pair shared between $A$ and $A^c$. This Bell pair contributes a fixed one bit to the entropy $S(A)$, independent of the remaining state.

With these simplifications, the remaining circuit defines an effective  \( [[8,2,3]] \) encoding circuit containing only the degrees of freedom that can affect the measured boundary entropy and the corresponding recovered bulk entropy. In this reduced description, the boundary subregion is denoted as $B$, and the effective minimal cut is reduced from three bonds in the original five-qubit boundary region to two bonds. Consequently, in the original stabilizer limit, the expected area-like contribution in this protocol is \(2\) bits.

\subsubsection{Magic injection}
In the $[[8,2,3]]$ encoding circuit, we offset all $R_x$ gates by $\lambda_x$, $R_y$ gates by $\lambda_y$, $R_z$ gates by $\lambda_z$, and $R_{XX}$ gates by $\lambda_{XX}$ as described in Appendix~\ref{app:sec:magic_injection}. Table~\ref{tab:lambdas_823} lists the gate type-dependent over-rotation parameters used for each value of total magic $M$.

\begin{table}[H]
\centering
\caption{Over-rotation parameters for the single $[[8,2,3]]$ code. Plot colors are a reference to Fig~\ref{fig:sc}. All angles are in radians.}
\label{tab:lambdas_823}
{
\setlength{\tabcolsep}{6pt} 
\begin{tabular}{cccccc}
\hline
 Plot color & $M$ & $\lambda_x$ & $\lambda_y$ & $\lambda_z$ & $\lambda_{XX}$ \\
\hline
\colorsquare{000000} & $0$    & $0$     & $0$     & $0$     & $0$    \\
\colorsquare{A395BF} & $2.23$ & $0.075$ & $0.125$ & $0.075$ & $0.10$ \\
\colorsquare{C6A9FF} & $5.79$ & $0.15$  & $0.25$  & $0.15$  & $0.20$ \\
\hline
\end{tabular}
}
\end{table}

\subsubsection{Recovery and measurement}

After preparing the reduced \( [[8,2,3]] \) encoded state, we apply the local recovery circuit and measure the boundary entropy $S_B$ and the entropy of the recovered bulk state $S_b$.

In the original HaPPY code,  the recovery circuit for the three-qubit subregion $B$ is given by the inverse of the corresponding green unitary block in the HaPPY encoding circuit, shown in Fig.~\ref{fig:happy25_circuit}. Therefore the recovery has a known local gate decomposition and exactly reconstructs the recoverable bulk degree of freedom from the chosen boundary region. When magic is injected into the encoding circuit, however, this Clifford recovery is no longer exactly optimal. We now keep the same recovery-circuit architecture but allow the individual gate parameters to vary, determining them by minimizing the cost function in Eq.~\eqref{eq:cost}. This gives an approximate recovery unitary adapted to the magic-enriched encoding while keeping the circuit depth experimentally feasible.

To measure the entanglement entropy, we perform state tomography on the three qubits in $B$, labeled $q_0,q_1,q_2$, after the recovery circuit on $B$ has been applied. The procedure is as follows:
\begin{enumerate}
    \item Measure the three-qubit Pauli string \(P_0 \otimes P_1 \otimes P_2 \) on qubits $q_0,q_1,q_2$,  and record the outcome as the bit-string \(s_0s_1s_2\). Note that the operators $P_0, P_1, P_2$ are all different from the identity. 
    
    \item Repeat the above steps for each of the $3^3$ Pauli string operators to reconstruct a 3-qubit reduced density matrix. Since the recovery circuit is unitary, it does not change the entropy of the reduced state. Thus, we can obtain the \textit{boundary} entropy $S_B$ from the full 3-qubit state, which we estimate with the reconstructed state after tomography.
    \item After applying the recovery unitary, the final state on the three qubits contains the logical information on one of the qubits (qubit $q_0$, say). By tracing out the other two qubits, we can thus compute the bulk entropy $S_b$ from $q_0$ alone.
    \item We perform this procedure for each selected value of the bulk entanglement, $\theta$. 
\end{enumerate}

We only sample $3^3$ weight-3 Pauli strings because the expectation value of lower weight Pauli string can be inferred from the data of weight-3 operators.

For both the boundary and recovered-bulk state tomography, we use maximum-likelihood estimation (MLE) \cite{PhysRevLett.108.070502} to obtain physical density matrices from the QST data.  

The proto-area entropy $S_{\text{PA}}$ is then obtained by subtracting the recovered-bulk entropy $S_b$ from the boundary entropy $S_B$.

\subsubsection{Entropy plots}
\begin{figure*}[h]
\centering
\includegraphics[width=\textwidth]{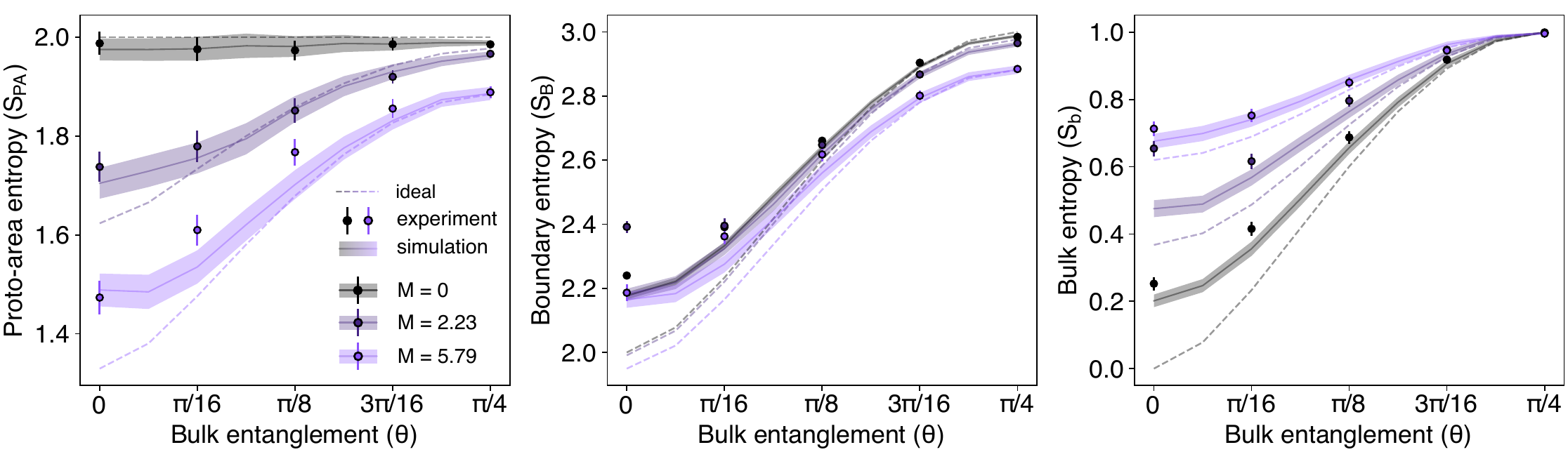}
\caption{
\textbf{Boundary and bulk entropy contributions to the single-code proto-area entropy.}
Extension of the single-code results shown in Fig.~\ref{fig:sc}b of the main manuscript.
The left panel shows the proto-area entropy $S_{\mathrm{PA}}=S_{\mathrm{B}}-S_{\mathrm{b}}$ as a function of the bulk entanglement angle $\theta$ for three values of injected magic.
The middle and right panels show the corresponding boundary entropy $S_{\mathrm{B}}$ and bulk entropy $S_{\mathrm{b}}$, respectively, from which $S_{\mathrm{PA}}$ is extracted.
Experimental points are obtained from quantum state tomography (QST), while shaded curves show noisy numerical simulations using the same analysis procedure. Ideal simulations assume no noise and perfect entropy reconstruction.
}
\label{fig:supp_sc8_plot}
\end{figure*}

The complete set of boundary and bulk entropies is shown in Fig.~\ref{fig:supp_sc8_plot}, including experimental data, noisy numerical simulations, and ideal or noiseless simulations. Error bars, obtained by bootstrapping, are computed separately for each subplot. Bootstrapping with replacement is used, with the number of repeats set equal to the number of shots of data in each case. Ideal simulations assume noiseless gates and perfect entropy reconstruction, and are calculated via exact statevector simulation of data circuits.

Overall, the ideal and noisy simulations agree well with the experimental data, as expected for relatively short circuits---$23$ $ZZ$ gates in this case, with noisy simulations assuming stochastic fluctuations in each gate angle (Appendix \ref{sec:noisynumerics}). As illustrated in Fig.~\ref{fig:supp_sc8_plot}, the deviation between the ideal simulation and the experimental data is most pronounced near $\theta=0$, where the measured entropy is small. This deviation is caused both by the hardware noise and finite shot effect in tomography. Importantly, these noise systematics tend to suppress the proto-area entropy signal, reducing the contrast between $\signal$ at $\theta=0$ and at $\theta=\pi/4$ relative to the ideal simulation. The experimentally measured $\signal$ should therefore be interpreted as a conservative estimate, or lower bound, of the signal induced by magic in the absence of stochastic noise.

\subsection{Two-sided $[[5,1,3]]$ code}\label{app:twosided}

We first implement a two-sided code circuit where each side is a \( [[5,1,3]] \) code. As shown in Fig.~\ref{fig:WHcircuit10}, the two logical qubits are initially entangled by the unitary \(U(\theta)\), which prepares the state 
\begin{equation}
    \cos\theta\,\ket{00}_{a_0b_0}+\sin\theta\,\ket{11}_{a_0b_0},
\end{equation}
where $a_0,b_0$ represent the bulk qubit in each side. 
Magic is injected into the encoding circuit by coherently over- or under-rotating the local gates in each encoding block, following the procedure described in Appendix~\ref{app:sec:magic_injection}.

The boundary region is chosen to be a 6-qubit combined subregion $C\cup C'$, consisting of three physical qubits from each code block. Without loss of generality, we take these to be the first three qubits of each block, as indicated in Fig.~\ref{fig:wh_data}c. 

\subsubsection{Encoding circuit}

\begin{figure}
    \centering
\includegraphics[width=0.8\linewidth]{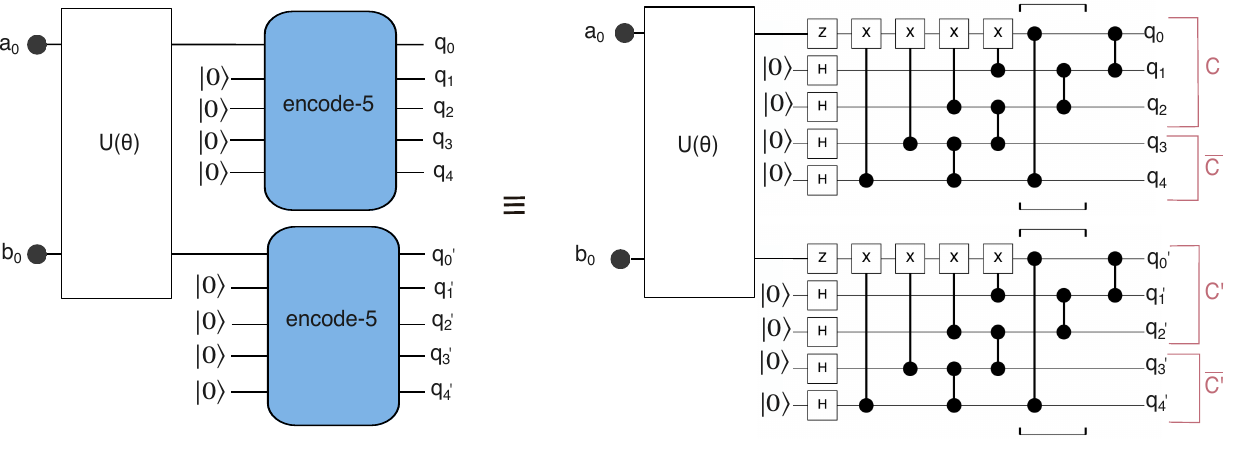}
    \caption{The encoding circuit for the 10-qubit two-sided code experiment.}
    \label{fig:WHcircuit10}
\end{figure}

The encoding circuit at zero-magic for the two-sided experiment is shown in Fig.~\ref{fig:WHcircuit10}. The entangled ``bulk'' state is encoded together with eight ancilla qubits initialized in \(\ket{0}\), producing a ten-qubit state across the two \( [[5,1,3]] \) code blocks. We label the five qubits in the first block by $q_0$--$q_4$ and qubits in the second block by $q_0'$--$q_4'$. 

\subsubsection{Magic injection} \label{app:sec:wh10_magic_injection}

The parameter \(\theta\in[0,\pi/4]\) controls the amount of  entanglement between the two bulk qubits. In the magic-enriched experiments, persistent coherent offsets \(\lambda_i\) are added to the relevant rotation angles, as described in Appendix~\ref{app:sec:magic_injection}. For experimental implementation, the Clifford encoding circuit is compiled into the native IonQ gate set. 

\begin{figure} 
    \centering
    \includegraphics[width=0.8\linewidth]{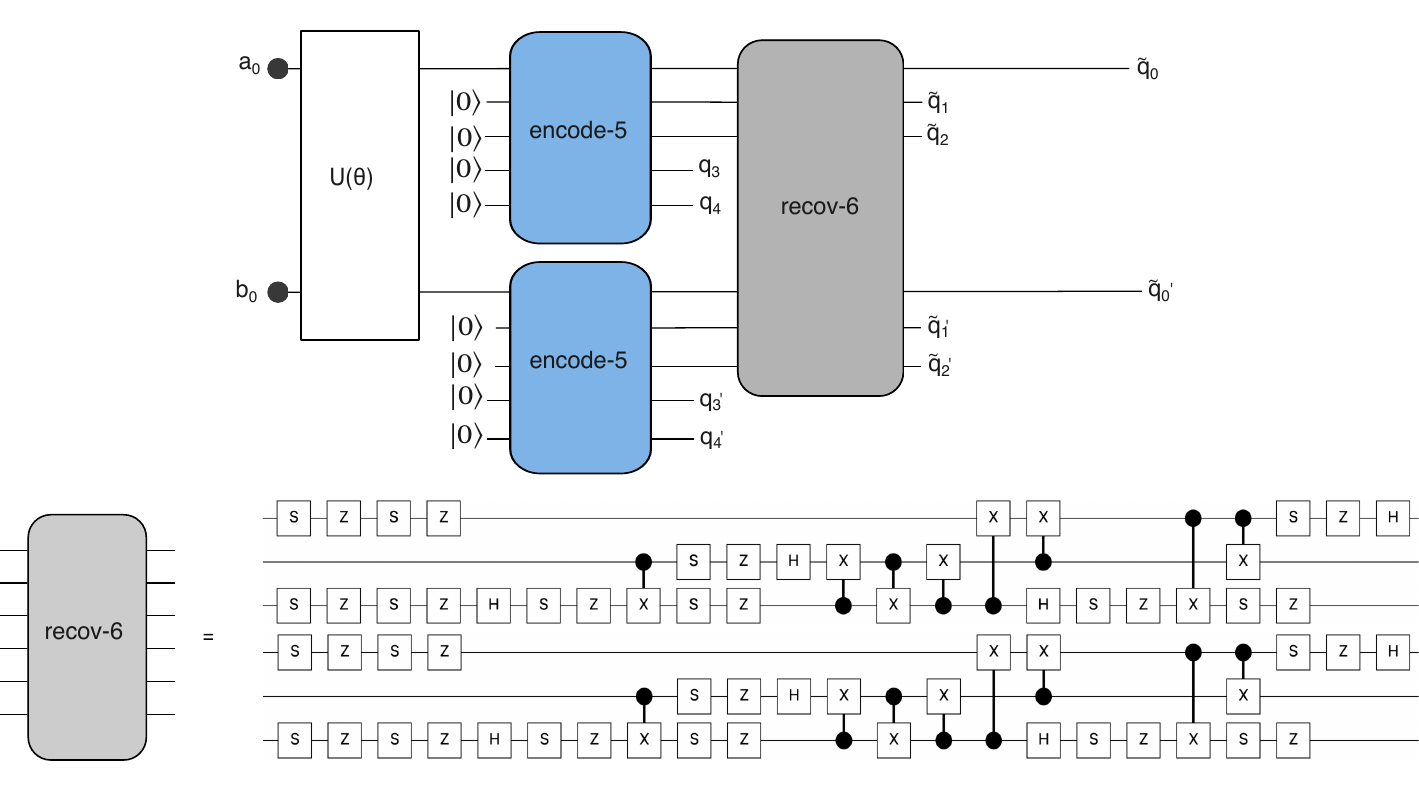}
    \caption{Clifford recovery circuit acting on the subregion $C\cup C'$, which contains six physical qubits $q_0$, $q_1$, $q_2$, $q_0'$, $q_1'$, $q_2'$. The recovered bulk state is supported on the output qubits $\tilde q_0, \tilde q_0'$. }    \label{fig:WHrecovery}
\end{figure}

Table~\ref{tab:lambdas_10wh} lists the gate type-dependent over-rotation parameters used for each value of total magic $M$.

\begin{table}[H]
\centering
\caption{Over-rotation parameters for the two-sided $[[5,1,3]]$ code. Plot colors are a reference to Fig~\ref{fig:wh_data}e. All angles are in radians.}
\label{tab:lambdas_10wh}
{
\setlength{\tabcolsep}{6pt} 
\begin{tabular}{cccccc}
\hline
Plot colors & $M$ & $\lambda_x$ & $\lambda_y$ & $\lambda_z$ & $\lambda_{XX}$ \\
\hline
\colorsquare{000000} & $0$    & $0$    & $0$    & $0$ & $0$     \\
\colorsquare{A395BF} & $1.51$ & $0.05$ & $0.10$ & $0$ & $0.135$ \\
\colorsquare{C6A9FF} & $4.63$ & $0.10$ & $0.20$ & $0$ & $0.27$  \\
\hline
\end{tabular}
}
\end{table}

\subsubsection{Recovery and measurement}\label{app:rectwoside5}
Similar to the single-copy case, we measure the boundary entropy  and the recovered bulk entropy  to determine $S_{\text{PA}}$.

The boundary entropy is given by the entropy of the 6-qubit subregion $C\cup C'$, where $C$ consists of $q_0$--$q_2$ and $C'$ consists of $q_0'$--$q_2'$ (see Fig.~\ref{fig:WHcircuit10}). Rather than performing tomography directly on this 6-qubit region, we measure its 4-qubit complement, consisting of the remaining two physical qubits from each code block--  $q_3$, $q_4$, $q_3'$, $q_4'$. Since the full ten-qubit encoded state is pure in the absence of hardware noise, the boundary entropy of $C\cup C'$ can be estimated from the entropy of its complement. We therefore obtain the boundary entropy from the four-qubit reduced density matrix on the complementary subregion. The tomography procedure is as follows: 

\begin{enumerate}
    \item Measure the four-qubit Pauli string 
    \(P_3\otimes P_4 \otimes P_3'\otimes P_4'\), and record the outcome as the bit-string \(s_3s_4s_3's_4'\).

    \item For each Pauli string, collect the corresponding experimental data. We use \(1250\)--\(1500\) shots per Pauli string.

    \item Repeat the above measurements for all \(3^4\) non-identity four-qubit Pauli strings to reconstruct the four-qubit reduced density matrix on the complementary boundary region. The reconstructed density matrix is then used to estimate the desired boundary entropy.

    \item Repeat this procedure for each selected value of the bulk entanglement angle \(\theta\).
\end{enumerate}

To obtain the recovered \textit{bulk} entropy, we apply the recovery unitary to the selected boundary region $C\cup C'$. We denote the output qubits of this recovery unitary by $\tilde q_0$-$\tilde q_2$ and $\tilde q_0'$-$\tilde q_2'$.  The resulting 6-qubit state contains logical information on the two output qubits $\tilde  q_0$ and $\tilde  q_0'$, which support the recovered bulk state. In the zero magic case, we use the (exact) Clifford recovery circuit illustrated in Fig.~\ref{fig:WHrecovery}, which consists of two copies of the recovery unitary used in the reduced $[[8,2,3]]$ code.  

When magic is injected, we minimize the cost function Eq.~\eqref{eq:cost} over the recovery circuit with the same architecture as the Clifford recovery circuit, allowing all the gate angles in the circuit to vary. 

After applying the optimal recovery circuit, the output qubits $\tilde  q_0$ and $\tilde  q_0'$ support the recovered bulk state (see Fig.~\ref{fig:WHrecovery}). We perform state tomography on these two qubits to obtain the recovered state, from which we compute the bulk entropy.  

For both the boundary and recovered-bulk state tomography, we use maximum-likelihood estimation (MLE) to obtain physical density matrices and suppress unphysical artifacts --- small negative eigenvalues from finite-shot statistical errors. 

The proto-area entropy is then obtained by subtracting the entropy of the recovered bulk state from the measured boundary entropy.
\begin{figure*}[t]
    \centering
    \includegraphics[width=\linewidth]{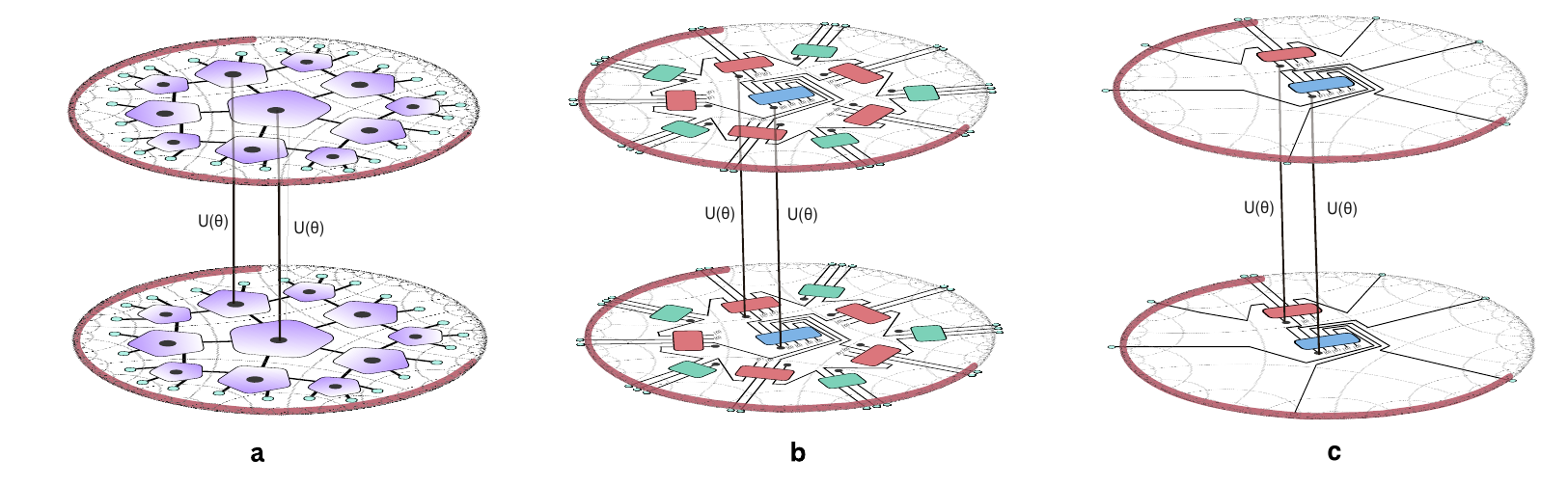}
    \caption{\textbf{Two-sided [[25,11,3]] HaPPY construction and its reduced circuit.}
\textbf{a}, Tensor-network representation of two coupled ( [[25,11,3]] ) HaPPY codes.
\textbf{b}, Corresponding two-sided circuit implementation.
\textbf{c}, Reduced 16-qubit circuit obtained from the isometric reduction, where the original 34-qubit boundary region (A) maps to the effective 12-qubit region (B).}
    \label{fig:twosided25}
\end{figure*}
\subsubsection{Entropy plots}
\begin{figure*}[t]
\centering
\includegraphics[width=\textwidth]{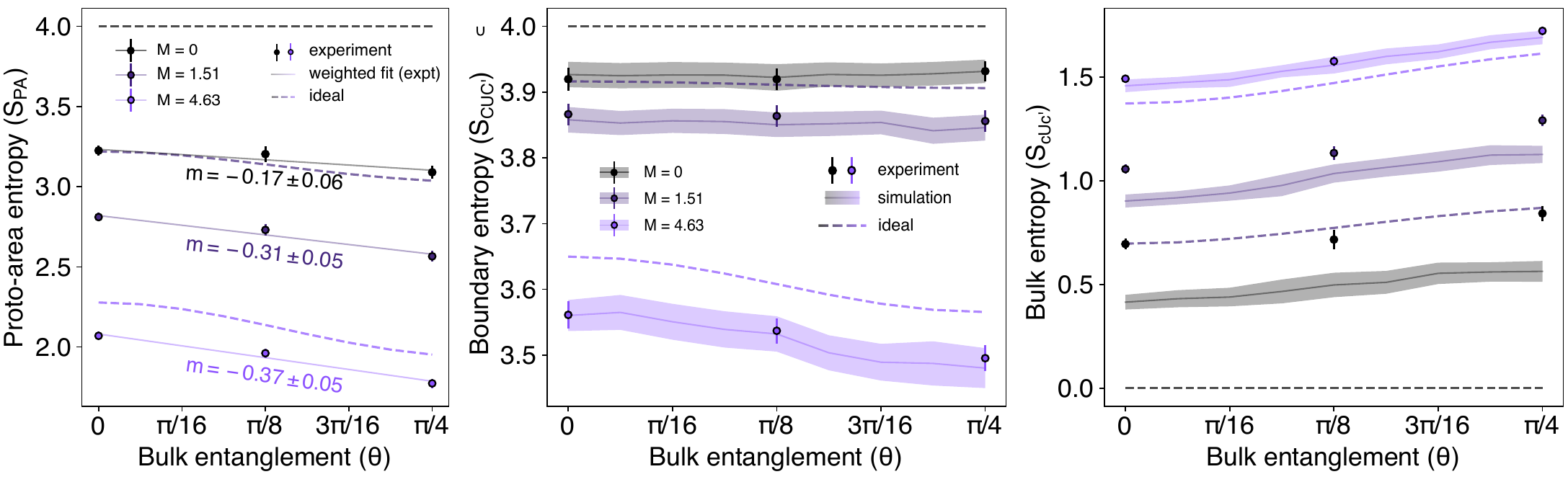}
\caption{
\textbf{Entropy components and fitted response for the  two-sided $[[5,1,3]]$ code experiment.}
Supplementary extension of Fig.~\ref{fig:wh_data}e in the main manuscript.
The left panel shows the proto-area entropy $S_{\mathrm{PA}}$ as a function of the bulk entanglement angle $\theta$ for the $2\times[[5,1,3]]$ code construction, showing experimental data and ideal noiseless calculations.  The experimental data is fitted with a weighted linear fit whose slope is printed here and whose magnitude is shown as a bar plot in Fig.~\ref{fig:wh_data}e.
The two right panels show the boundary entropy $S_{\mathrm{bound}}$ and bulk entropy $S_{\mathrm{bulk}}$ used to compute $S_{\mathrm{PA}}=S_{\mathrm{bound}}-S_{\mathrm{bulk}}$.
Experimental entropies are estimated by MLE from quantum state tomography, with error bars denoting $\pm1\sigma$ bootstrap uncertainties; shaded bands show the corresponding noisy simulation results.
}
\label{fig:supp_wh10_plot}
\end{figure*}

The complete set of experimentally measured entropies for the 10-qubit two-sided experiment is shown in Fig.~\ref{fig:supp_wh10_plot}. The noisy numerical simulations agree well with the experimental data for both the boundary entropy and the recovered bulk entropy.

The simulations include only stochastic fluctuations in the gate angles, whose values are listed in Table~\ref{tab:calculatd_sigma_mu}. The zero-magic experiments, $M=0$, are particularly sensitive to these errors, which appear as an approximately vertical offset in the entropy-versus-$\theta$ curves. Adding fixed gate-angle offsets, which model coherent errors, reduces the discrepancy between experiment and noisy simulations, as shown in Figure~\ref{fig:wh10_fitted_qst_noise}.

\subsection{Two-sided $[[8,2,3]]$ code}

We now extend the two-sided construction to the case in which each side is a 25-qubit HaPPY code. As before, the two copies are coupled by entangling their bulk qubits, as shown in Fig.~\ref{fig:twosided25}a.

To probe the connectivity of the emergent geometry, we measure the proto-area entropy $S_{\text{PA}}$ for a combined boundary region formed from two disjoint 17-qubit subregions, one in each copy of the code; see Fig.~\ref{fig:twosided25}a. Following the same reduction procedure as in Appendix~\ref{app:reduction}, each 25-qubit code can be reduced to an effective $[[8,2,3]]$ code. Consequently, the original 34-qubit boundary region is mapped to the 12-qubit region $D\cup D'$ in the two-sided reduced code, as shown in Fig.~\ref{fig:twosided25}c.

\subsubsection{Encoding circuit}

\begin{figure*}[t]
    \centering
    \includegraphics[width=0.8\linewidth]{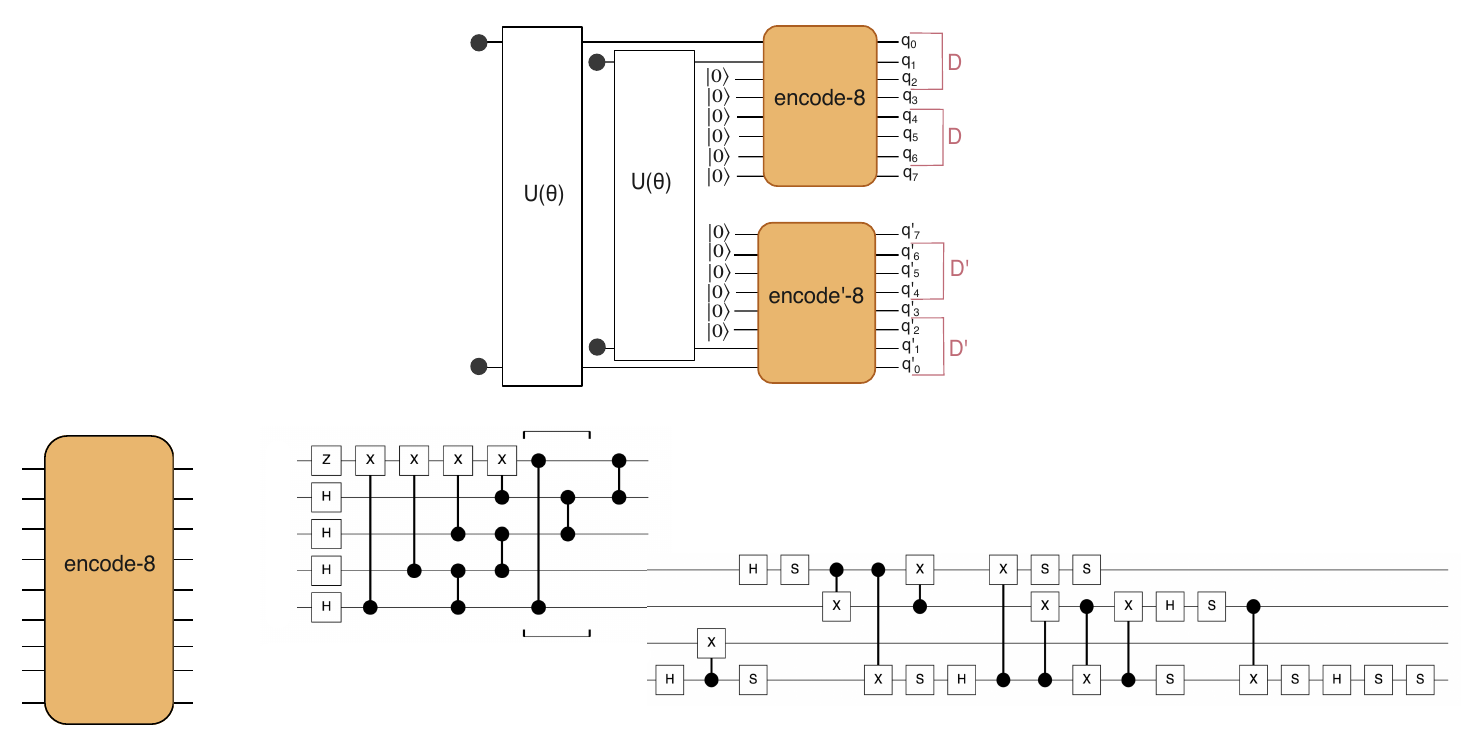}
    \caption{ The encoding circuit for the two-sided [[8,2,3]] code experiment. Here, the $\mathrm{encode}'$-8 circuit denotes the same circuit as $\mathrm{encode}$-8, but with the ordering of both the input and output qubits reversed. }
    \label{fig:wh16_circuit}
\end{figure*}

The encoding circuit for the two-sided reduced HaPPY code consists of two copies of the $[[8,2,3]]$ encoding circuit. In the zero-magic limit, its gate decomposition is shown in Fig.~\ref{fig:wh16_circuit}. The four input bulk qubits are grouped into two pairs, each entangled by a unitary $U(\theta)$ with the same parameter $\theta\in[0,\pi/4]$. This parameter controls the amount of bulk entanglement in each pair, preparing the state
\begin{equation}\small\left(\cos\theta\,\ket{00}_{a_0b_0}+\sin\theta\,\ket{11}_{a_0b_0}\right)\otimes \left(\cos\theta\,\ket{00}_{a_1b_1}+\sin\theta\,\ket{11}_{a_1b_1}\right).
\end{equation}
where $a_0,a_1$ and $b_0,b_1$ denote the bulk qubits in the first and second copies of the $[[8,2,3]]$ code, respectively.

As shown in Fig.~\ref{fig:wh16_circuit}, the circuit has 16 output qubits, labelled $q_0$--$q_7$ and $q'_0$--$q'_7$ for the two blocks, respectively. The subregion $D$ consists of the qubits $q_0$--$q_2$ together with $q_4$--$q_6$, while $D'$ consists of the corresponding qubits $q'_0$--$q'_2$ and $q'_4$--$q'_6$ in the second block.
\subsubsection{Magic injection} \label{app:sec:wh16_magic_injection}

For the two-sided $[[8,2,3]]$, the magic injection is performed using gate-specific perturbations. Each single- and two-qubit rotation angle in the encoding circuit was shifted by an independently optimized offset $\lambda_i$, where $i$ labels the corresponding gate in the circuit. The offsets are constrained to $\lambda_i \in [-0.1,0.1]$~radians. The optimized set of offsets is chosen to enhance the \signal~signal contrast between $\theta=0$ and $\theta=\pi/4$. 

\begin{figure*}[t]
    \centering
    \includegraphics[width=0.8\linewidth]{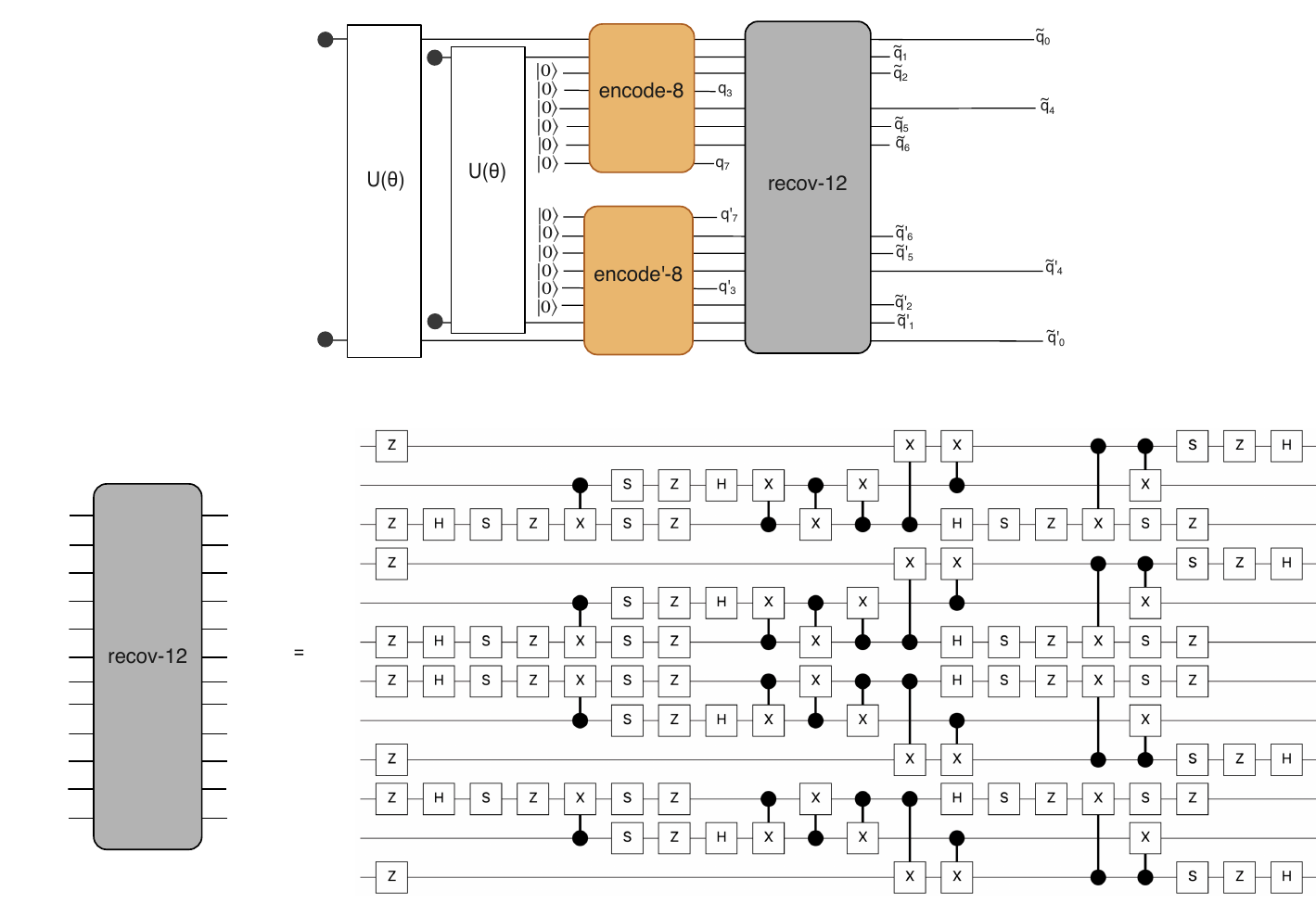}
    \caption{The Clifford recovery circuit applied to the subregion $D\cup D'$, which contains physical qubits $q_0$-$q_2$, $q_4$-$q_6$, $q_0'$-$q_2'$, and $q_4'$-$q_6'$.  The recovered bulk state is supported on the output qubits $\tilde q_0, \tilde q_4, \tilde q_4', \tilde q_0'$. }
    \label{fig:wh16_recover}
\end{figure*}

\subsubsection{Recovery and measurement }

To measure the boundary entropy, we perform quantum state tomography on the complement of the subregion $D\cup D'$, which consists of the four qubits $q_3,q_7,q'_3,q'_7$, as shown in Fig.~\ref{fig:wh16_circuit}. Since the ideal global state is pure (i.e., in the absence of hardware noise), the entropy of this four-qubit complement provides an estimate of the boundary entropy of $D\cup D'$.

The entropy-measurement procedure follows that described in Appendix~\ref{app:rectwoside5}. We measure the $3^4$ strings of Pauli operators of the form $P_3\otimes P_7\otimes P'_3\otimes P'_7$. Each  Pauli string is measured using 1000 shots to estimate its expectation value.

In the zero-magic case, the bulk state is recovered by applying two copies of the local Clifford recovery circuit to the subregion $D\cup D'$, as shown in Fig.~\ref{fig:wh16_recover}. The recovered bulk state is supported on the four qubits $\tilde q_0,\tilde q_4,\tilde q'_0,\tilde q'_4$. When magic is injected, we instead optimize the recovery unitary while keeping the same circuit architecture as the Clifford recovery.

To obtain the bulk entropy, we perform tomography on the \textit{recovered} four-qubit bulk state supported on $\tilde q_0,\tilde q_4,\tilde q'_0,\tilde q'_4$ after applying the recovery circuit. Specifically, we measure the $3^4$ Pauli-string operators of the form $P_0\otimes P_4\otimes P'_0\otimes P'_4$, again multiple times (shots) for each operator. 

\subsubsection{Entropy plots}

\begin{figure*}[b]
\centering
\includegraphics[width=\textwidth]{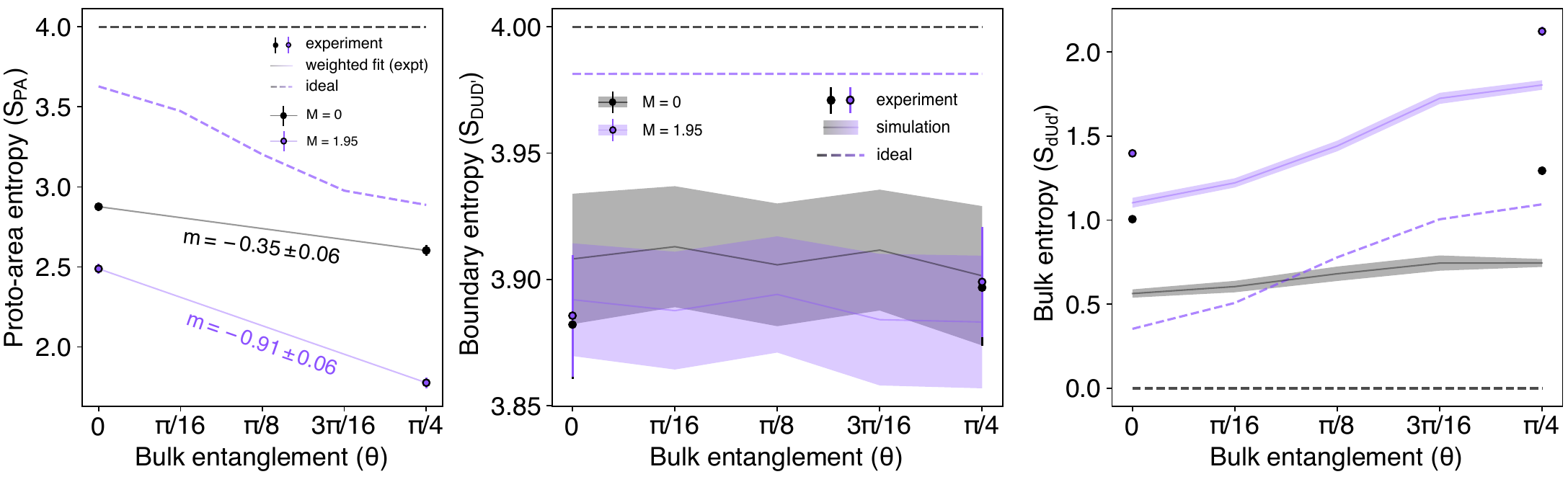}
\caption{
\textbf{Entropy components and fitted response for the  two-sided $[[8,2,3]]$ code experiment.}
Supplementary extension of Fig.~\ref{fig:wh_data}f in the main manuscript.
The left panel shows the proto-area entropy $S_{\mathrm{PA}}$ as a function of the bulk entanglement angle $\theta$ for the $2\times[[8,2,3]]$ code construction, showing experimental data and ideal noiseless calculations. The experimental data is fitted with a weighted linear fit whose slope is printed here and whose magnitude is shown as a bar plot in Fig.~\ref{fig:wh_data}f.
The two right panels show the boundary entropy $S_{\mathrm{bound}}$ and bulk entropy $S_{\mathrm{bulk}}$ used to compute $S_{\mathrm{PA}}=S_{\mathrm{bound}}-S_{\mathrm{bulk}}$.
Experimental entropies are estimated by MLE from quantum state tomography, with error bars denoting $\pm1\sigma$ bootstrap uncertainties; shaded bands show the corresponding noisy simulation results.
}
\label{fig:supp_wh16_plot}
\end{figure*}

Figure~\ref{fig:supp_wh16_plot} shows the experimental data and numerical simulations for the $2\times[[8,2,3]]$ code experiment. The experiment and noisy simulation agree well in boundary entropy experiment but they are off by a vertical offset in the bulk entropy case. This discrepancy is larger in the $M=0$ case, whose ideal and simulated values are already very small. This mismatch is likely amplified by errors in density-matrix and entropy reconstruction, which become more pronounced when the target bulk entanglement is low and the number of measured qubits is large (4 qubits in this case).

The noisy simulations do not include coherent gate-angle errors, which could explain the large discrepancy in these deeper circuits. In Fig.~\ref{fig:wh16_fitted_qst_noise} we show that fitting coherent error parameters $\mu_{1q},\mu_{2q}$ to the experimental data could reduce the discrepancy between experiment and simulation. Increasing the coherent offsets shifts the signal in both the $M=0$ and $M=1.95$ cases, but does not totally explain the large discrepancy observed, especially in the $M=0$ bulk entropy.

\section{Noise Analysis}
\label{sec:noisynumerics}
This section analyzes the effect of hardware noise on our experiments. 
We assume a simple noise model where the single- and two-qubit gates have stochastic fluctuations around their intended gate angles, whose magnitude is informed by independent gate fidelity measurements we interspersed during our data collection. The dominant noise in the Forte-1 system is reportedly stochastic in nature~\cite{forte_benchmarking}. Coherent errors (fixed gate angle offsets error) are also likely present, but a rigorous quantitative estimate has not been possible with the fidelity measurements at hand (\ref{app:noise_stochastic_errors}).

Other noise sources, including State Preparation and Measurement (SPAM) error and crosstalk, are not explicitly modeled: SPAM errors occur only once per circuit and are benchmarked at approximately $0.5\%$ per qubit~\cite{forte_benchmarking}, while crosstalk effects are expected to be partially captured by the representative gate-fidelity checks.

To compare to experimental data, we used simulations of circuits with noisy gate angles and emulated state tomography and entropy reconstruction with the same number of shots as in the experiment. This ensures that finite-shot effects and reconstruction bias coming from MLE are replicated in the simulation as much as possible.
Stochastic errors are modeled as gate-angle fluctuations which are resampled independently for every gate and shot from a zero-mean Gaussian with standard deviation $\sigma$. The specific choices of $\sigma$ in each experiment are based on fidelity measurements during the experimental run.

Sec~\ref{app:noise_stochastic_errors} explains gate fidelity measurements on the trapped-ion quantum computer used to estimate the amount of stochastic errors $\sigma$. Sec~\ref{app:noise_coherent_fit} compares noisy numerical simulations with the experimental data and analyzes coherent gate-angle offsets as a possible source of the remaining discrepancy. 
\ref{app:noisysystematics} describes the effects of stochastic noise on the proto-area entropy and explains the origin of the nonzero slope observed in the zero-magic two-sided experiment.

\subsection{Stochastic errors from gate fidelity measurements} \label{app:noise_stochastic_errors}

Errors in quantum hardware can be broadly classified into coherent and stochastic errors. Coherent errors are persistent, unitary errors in which the implemented operation differs from the target operation by an additional unitary rotation. They can accumulate constructively over repeated gate applications. Stochastic errors, by contrast, are an irreversible noise process that cannot be represented as a fixed unitary rotation. Both types of errors affect the measured entropy, so we attempt to quantify them using gate fidelity measurements we run on the quantum hardware. However, our checks did not give us a reliable measure of the coherent errors. So, we used them as an estimator of stochastic errors only, as explained below.

We first check the single-qubit errors by applying up to $10^3$ repeated GPi$(0)$ gates, equivalent to $R_x(\pi)$, and fitting the decay of the expected population of $\lvert 1\rangle$ state. From three randomly chosen ion qubits, we extract a weighted mean single-qubit gate fidelity of $0.999954(3)$. Since this error rate is much smaller than two-qubit gate errors, we focus the remaining diagnostics on the two-qubit gates.

We characterize two-qubit gate performance by preparing a Bell state from $\lvert 00\rangle$ using an implemented effective $R_{XX}(\pi/2)$ gate, following the transpilation procedure described in Sec.~\ref{sec:transpilation}. The Bell-state fidelity is estimated from the even-parity populations and the parity-fringe contrast~\cite{ballance2014high},
\begin{equation}
\mathcal{F}_{2q}
=
\frac{1}{2}
\left(
P_{00}
+
P_{11}
+
C_\Pi
\right),
\end{equation}
where $P_{00}$ and $P_{11}$ are measured immediately after Bell-state preparation. The parity-fringe contrast $C_\Pi$ is extracted from the phase-dependent parity
$$
\Pi(\phi)
=
P_{00}(\phi)
+
P_{11}(\phi)
-
P_{01}(\phi)
-
P_{10}(\phi),
$$
measured after applying analysis pulses with phase $\phi$. In our implementation, we estimate $C_\Pi$ from two analyzer phases, $\phi=\pi/4$ and $3\pi/4$, after verifying the expected parity-fringe phase. We repeat this procedure for sequences containing an odd number of effective two-qubit gates and extract the per-gate fidelity from a linear fit to the Bell-state fidelity versus gate count, as shown in Fig.~\ref{fig:cumulative_echoed_spot_check_q13_23}.

\subsubsection{Fidelity spot checks}

During each experiment, we monitor representative two-qubit gate pairs that occur frequently in the compiled circuits. These fidelity spot checks are used to estimate the gate-error parameters entering the noisy entropy simulations and are performed using the following two sequences.

\textbf{Cumulative spot checks:} these are designed to accumulate errors over repeated entangling gates.
We initialize both qubits in $|00\rangle$ and run repeated $ZZ$ gates $2N+1$ times, with the sequence wrapped by single-qubit $\pi/2$ rotations. 

The sequence is therefore 
$$\ket{00} \text{---} R_y^{\otimes2}(\pi/2) \text{---} [ZZ(\vartheta)]^{2N+1} \text{---} R_y^{\otimes2}(-\pi/2)\text{---}R_\phi(\pi/2) \text{---} M$$

\textbf{Echoed spot checks.} These sequences are designed to suppress fixed coherent errors by alternating the sign of the native entangling gate. Starting from $\lvert 00\rangle$, we apply the sequence

$$\ket{00} \text{---} R_y^{\otimes2}(\pi/2) \text{---} \left[ZZ(\vartheta)ZZ(-\vartheta)\right]^N \text{---} ZZ(\vartheta) \text{---} R_y^{\otimes2}(-\pi/2)\text{---}R_\phi(\pi/2) \text{---} M$$

with the same single-qubit rotations and analysis pulses used in the cumulative spot checks.

In principle, the echoed and cumulative spot checks can be used to separate stochastic and coherent error contributions: the echoed sequence should suppress fixed coherent over-rotations, while the cumulative sequence should amplify them. Echoed gate fidelity usually scales linearly with the number of gates. Cumulative gate fidelity, on the other hand, can have a quadratically decaying term as well due to the coherent buildup of errors over successive gates, and should have higher error rates \cite{lairdeganthesis}. In our measurements, however, the echoed sequences produce larger error rates than the cumulative sequences, so this separation is not reliable. At the same time, the measured cumulative gate fidelities reflect an approximately linear dependence on the number of gates, suggesting that coherent errors affecting the qubit pair are negligible compared to stochastic errors. We therefore use the cumulative spot-check results to set a conservative stochastic gate-error scale in the noisy simulations, and estimate coherent offsets separately by matching to experimental data.

\subsubsection{Implementation}
    \begin{figure}[htbp]
    \centering
    \includegraphics[width=0.5\textwidth,trim=290 0 0 0,clip]{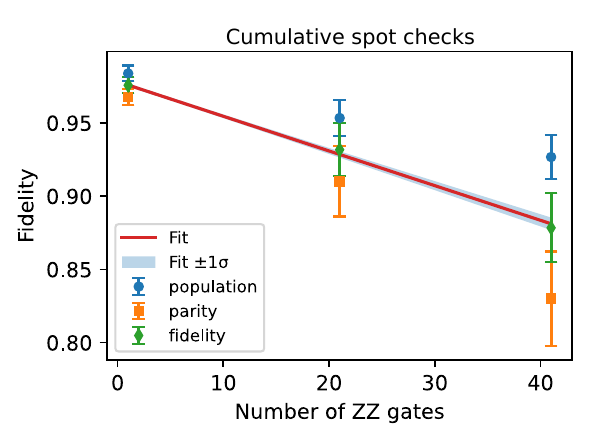}
    \caption{Example fidelity spot check for the cumulative \(ZZ\)-gate characterization sequence on qubit pair \((13,23)\). The plot shows fidelity, population, and parity-coherence contributions for increasing number of applied \(ZZ\) gates. A linear fit to the extracted fidelity is shown with a \(1\sigma\) confidence band, giving an estimated per-gate infidelity of \(1-F_g \approx 0.0024 \pm 0.0001\).}    \label{fig:cumulative_echoed_spot_check_q13_23}
\end{figure}

The circuits were run on IonQ Forte-1~\cite{forte_benchmarking} during $8-15$~hour windows of exclusive runtime access to the machine. 
This ensured data collection with minimal calibration drift.
At the start of each session, we select an optimized qubit-to-ion mapping using the latest IonQ calibration data. We then perform spot checks of the relevant ion-pair fidelities and revise the mapping if any measured fidelities fall below an acceptable threshold. Once a suitable mapping is identified, it is held fixed unless later checks reveal significant calibration drift. To monitor changes in two-qubit error rates throughout the experiment, we interleave fidelity spot checks with batches of data circuits, typically at approximately hourly intervals. Our custom software toolkit coordinates the submission and analysis of data and calibration circuits in close succession, thereby minimizing the effects of drift between the corresponding measurements \cite{ionq-experiment-toolkit}. 
The average weighted fidelity for each experiment is obtained by first averaging the measured fidelity spot checks for each map, for both echoed and the cumulative versions, and then calculating the weighted mean and standard deviations, where the weight is the fraction of shots measured with a certain map. These values are presented in table \ref{tab:fidelity_mean_std}

\begin{table}[htbp]
    \centering
    \begin{tabular}{lcc}
        \hline
        Experiment & 
        \(\overline{F}_{\mathrm{cumul.}}\) & 
        \(\sigma_{\mathrm{cumul.}}\) \\
        \hline
        Single-copy $[[8,2,3]]$ Experiment & 0.9960 & 0.0015 \\
        Two-sided $[[5,1,3]]$ Experiment  & 0.9966 & 0.0010 \\        
        Two-sided $[[8,2,3]]$ Experiment  & 0.9961 & 0.0014 \\
        \hline
    \end{tabular}
    \caption{Weighted mean and weighted standard deviation of cumulative \(ZZ\)-gate fidelities for each experiment. The proportion of shots used with each applied qubit mapping is used as the weight.}
    \label{tab:fidelity_mean_std}
\end{table}

As noted above, the echoed sequences produce lower fidelities than the cumulative sequences on Forte-1, contrary to the expectation that echoing should suppress fixed coherent errors. 
Possible reasons for this include: colored or non-Markovian noise; errors due to periodic or non-stochastic noise that do not cancel out when echoed; and technical coherent noise that affects specific sub-components of the $ZZ$ gate in Forte, which is constructed as a composite gate with additional single-qubit rotations~\cite{forte_benchmarking}. Identifying the underlying mechanism however is beyond the scope of this work.

We assume the loss in fidelity of gates is dominated by stochastic errors---based on the linear relationship we observe in Fig~\ref{fig:cumulative_echoed_spot_check_q13_23}b \cite{lairdeganthesis} and on IonQ's previous benchmarking \cite{forte_benchmarking}. We model these stochastic errors as randomly fluctuating gate angles in our simulations. 
For example, this would change $ZZ_i(\pi/2) \longrightarrow ZZ_i(\pi/2+\epsilon_i)$, where $i$ is an index for the specific gate and  $\epsilon_i$ is the angle error in radians.
This results in a reduced two-qubit gate fidelity given by $\mathcal{F}(\epsilon) = \cos^2\left(\frac{\epsilon}{2}\right)$,
where $\mathcal{F}(\epsilon)$ is the fidelity as a function of the gate angle error $\epsilon$ in radians. This gives an estimate of the stochastic gate-angle spread. For example, a fidelity of $\mathcal{F}=0.9955$ corresponds to $\epsilon \simeq 0.135~\mathrm{rad}$. The calculated values of $\sigma$ that are used in the simulation are tabulated in \ref{tab:calculatd_sigma_mu}. 

Our noisy numerical simulations inject these errors into the native-gate layer to model noise on the hardware. For each shot and each gate, we allow the relevant native-gate angle to vary by an amount $\epsilon_i$, where $\epsilon_i$ is drawn from a Gaussian distribution with a spread equal to the calculated $\sigma$ values in Table~\ref{tab:calculatd_sigma_mu}. IonQ uses single-qubit gates $\text{GPi}(\varphi)$ and $\text{GPi2}(\varphi)$, where $\varphi$ defines the axis of rotation for $\pi$- and $\pi/2$-pulses, respectively. Although $\varphi$ is not the rotation angle itself, we stochastically vary this parameter as an effective model for single-qubit gate noise on the hardware. The native two-qubit gates are $ZZ(\vartheta)$ gates, and we stochastically vary $\vartheta$. In all noisy simulations, the stochastic gate-angle fluctuations are resampled independently for every gate and shot from a zero-mean Gaussian with standard deviation $\sigma$.

\subsection{Coherent errors estimated from data circuits} \label{app:noise_coherent_fit}

\begin{table}[htbp]
    \centering
    \begin{tabular}{l@{\hspace{1.0cm}}c@{\hspace{0.8cm}}c@{\hspace{0.8cm}}c@{\hspace{0.8cm}}c}
        \hline
        & \multicolumn{2}{c@{\hspace{0.8cm}}}{Fidelity spot checks}
        & \multicolumn{2}{c}{Fit to experimental data} \\
        \cline{2-5}
        Experiment &
        $\sigma_{\mathrm{1q}}$ &
        $\sigma_{\mathrm{2q}}$ &
        $\mu_{\mathrm{1q}}$ &
        $\mu_{\mathrm{2q}}$ \\
        \hline
        Single-copy $[[8,2,3]]$ code & 0.02 & $\sigmaZZsc(24)$ & -- & -- \\
        Two-sided $[[5,1,3]]$ code & 0.02 & $\sigmaZZwh(17)$ & 0.01 & 0.068(63) \\
        Two-sided $[[8,2,3]]$ code & 0.02 & $\sigmaZZdwh(23)$ & 0.01 & 0.046(1) \\
        \hline
    \end{tabular}
    \caption{Noise parameters used in noisy numerical simulations. The stochastic error widths, $\sigma_{\mathrm{1q}}$ and $\sigma_{\mathrm{2q}}$, are derived from fidelity spot checks and are used as the baseline noise parameters. The coherent offsets, $\mu_{\mathrm{1q}}$ and $\mu_{\mathrm{2q}}$, are obtained from fits to experimental entropy data and are shown as optional fitted parameters; their uncertainties are large because the fitted slopes have substantial uncertainty. All units are radians.}
    \label{tab:calculatd_sigma_mu}
\end{table}

\begin{figure*}[t]
    \centering
    \includegraphics[width=\textwidth]{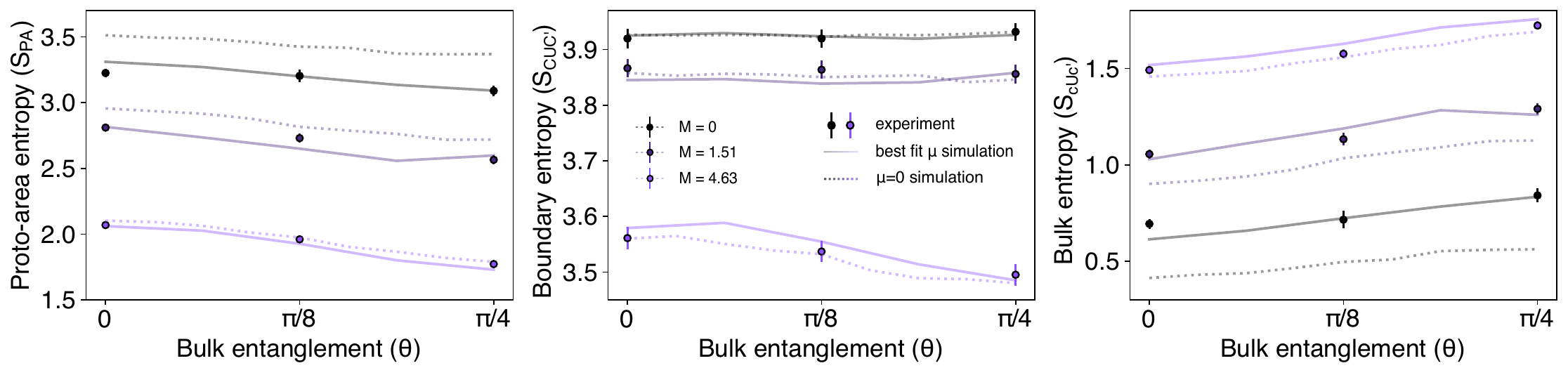}
    \caption{
    \textbf{Noise-model comparison for the two-sided $[[5,1,3]]$ experiment.}
    Proto-area entropy $S_{\mathrm{PA}}$ (left), boundary entropy $S_{\mathrm{C \cup C^\prime}}$ (middle), and bulk entropy $S_{\mathrm{c \cup c^\prime}}$ (right) are plotted as functions of the input bulk entanglement angle $\theta$ for three magic parameters. Filled circles with error bars show the experimental data. Dotted curves show noisy-QST simulations with only stochastic noise, as used in the main text, while solid curves show fitted noisy simulations that include small coherent gate-angle offsets. The best-fit coherent $ZZ$ error is $\mu_{ZZ} = 0.068(63)$. We set $\mu_{1q} = 0.2\times\mu_{ZZ}$, giving $\mu_{1q} \approx 0.01$. Including coherent offsets improves the agreement between the noisy simulations and the observed entropy trends. 
    }
    \label{fig:wh10_fitted_qst_noise}
\end{figure*}

For deeper circuits ($\gtrsim20$ $ZZ$ gates) like in the two-sided experiments, the stochastic unitary noise model alone does not perfectly match the experimental data, as shown in Fig.~\ref{fig:wh_data}, particularly in the no-injected-magic cases ($M=0$).  The possible reason for this mismatch is from the coherent gate-angle offsets, which can accumulate systematically in deeper circuits. Physically, this would come from gate miscalibration, leading to fixed angle offset. 

\begin{figure*}[t]
    \centering
    \includegraphics[width=\textwidth]{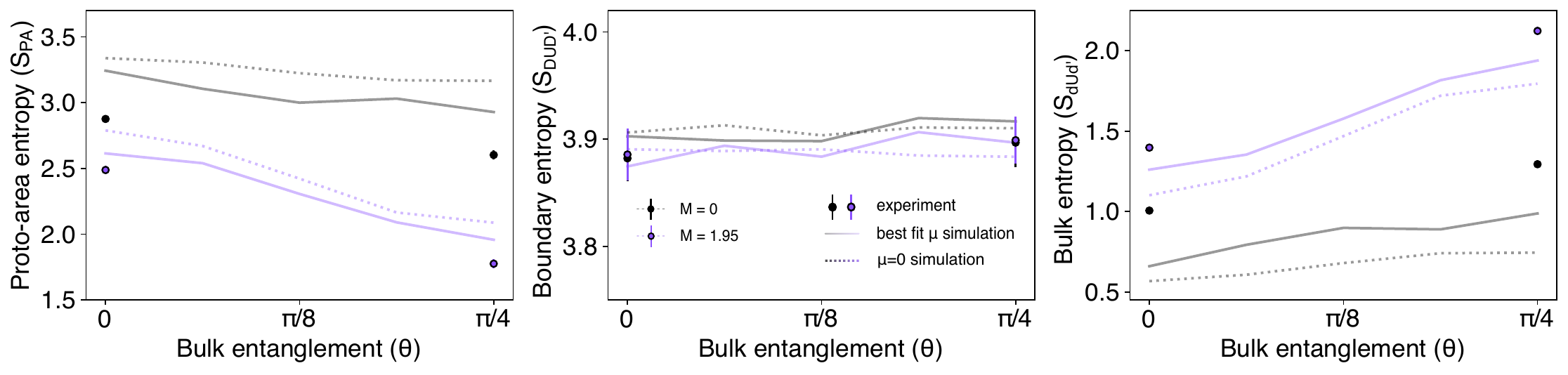}
    \caption{
    \textbf{Noise-model comparison for the two-sided $[[8,2,3]]$ experiment.}
    Proto-area entropy $S_{\mathrm{PA}}$ (left), boundary entropy $S_{\mathrm{D \cup D^\prime}}$ (middle), and bulk entropy $S_{\mathrm{d \cup d^\prime}}$ (right) are plotted as functions of the input bulk entanglement angle $\theta$ for two magic parameters. Filled circles with error bars show the experimental data. Dotted curves show noisy-QST simulations with only stochastic noise, as used in the main text, while solid curves show fitted noisy simulations that include small coherent gate-angle offsets. The best-fit coherent $ZZ$ error is $\mu_{ZZ} = 0.046(1)$. We set $\mu_{1q} = 0.2\times\mu_{ZZ}$, giving $\mu_{1q} \approx 0.01$. Including coherent offsets improves the agreement between the noisy simulations and the observed entropy trends. 
    }
    \label{fig:wh16_fitted_qst_noise}
\end{figure*}

To estimate the scale of coherent errors needed to account for the remaining discrepancy, we analyze the two-sided $[[5,1,3]]$ experiment shown in Fig.~\ref{fig:wh10_fitted_qst_noise}. In this data set, there is a noticeable mismatch between the experimental data and noisy simulations  that include only stochastic errors, corresponding to $\mu=0$ and shown by the dotted line. In these simulations, the stochastic error strengths are fixed to \(\sigma_{ZZ}=\sigmaZZwh~\mathrm{rad}\) and \(\sigma_{\mathrm{1q}}=0.02~\mathrm{rad}\). 
We then fit noisy QST simulations to the bulk and boundary entropy data points from the two-sided $[[5,1,3]]$ experiment, varying the two-qubit coherent offset error \(\mu_{\mathrm{2q}}\). The single-qubit coherent offset error is tied to this value by \(\mu_{\mathrm{1q}}=0.2\mu_{\mathrm{2q}}\), with the ratio chosen to roughly match the measured ratio of \(\sigma_{\mathrm{1q}}\) to \(\sigma_{ZZ}\). This procedure gives a best-fit value \(\mu_{\mathrm{2q}}=0.068\pm0.063~\mathrm{rad}\)  and \(\mu_{\mathrm{1q}}=0.016\pm0.013~\mathrm{rad}\). The improved agreement suggests that small coherent gate-angle offset errors are a plausible contribution to the residual mismatch between the stochastic-only simulation and the experimental data. A similar analysis is performed for the two-sided $[[8,2,3]]$ experiment, with the corresponding entropy plots shown in Fig.~\ref{fig:wh16_fitted_qst_noise}.

The uncertainty in the fitted $\mu_{\mathrm{2q}}$ is estimated by propagating the combined uncertainty of the experimental and noisy-simulation slopes through the fitted relation between the simulated slope and $\mu_{\mathrm{2q}}$. This uncertainty is large because the slope is only weakly sensitive to changes in $\mu$. Moreover, the cumulative-fidelity measurements in \ref{app:noise_stochastic_errors} indicate that the coherent errors are significantly smaller than stochastic errors. Therefore, the actual coherent error on the hardware is likely smaller than the best-fit value above, which is already a substantial fraction of the stochastic two-qubit error scale \(\sigma_{ZZ}\approx0.135~\mathrm{rad}\).

\subsection{Noisy Systematics} \label{app:noisysystematics}

In this section, we further analyze the different ways in which coherent and stochastic noise affect the proto-area entropy, and demonstrate a strategy for cancelling the leading contribution from stochastic noise.  
Since our main diagnostic is the bulk-state dependence of the proto-area entropy, noise is relevant only if it induces a $\theta$-dependent shift in $S_{\mathrm{PA}}$.

As explained in Appendix~\ref{app:noise_stochastic_errors}, coherent errors act as fixed perturbations  and can be viewed as injecting additional magic into the circuit. They can therefore directly modify the $\theta$ dependence of the proto-area entropy. In particular, increasing the coherent error amplitude generally increases the magnitude of the observed slope in $S_{\mathrm{PA}}(\theta)$.

For the stochastic noise,  we model it as random fluctuations in local gate angles. After averaging over repeated circuit executions, these fluctuations act approximately as local depolarizing noise. Schematically, for a random Hermitian generator $W$ and a small random angle $\epsilon$, the averaged channel takes the form
\begin{equation}
\mathbb{E}_{W}
\left[
e^{i\epsilon W}\rho e^{-i\epsilon W}
\right]
\simeq
(1-p_{\epsilon})\rho
+
p_{\epsilon}\frac{I}{d},
\end{equation}
where $p_{\epsilon}$ is set by the variance of the angle fluctuations and $d$ is the Hilbert-space dimension of the local subsystem.

Depolarizing noise can also generate $\theta$-dependent shifts in both the boundary entropy and the recovered bulk entropy. A simple example is a two-qubit bulk state with entanglement controlled by $\theta$. Local depolarizing noise makes this state mixed, producing a nonzero entropy contribution whose magnitude is positively correlated with the initial entanglement. This is consistent with the zero-magic bulk-entropy data shown in Fig.~\ref{fig:supp_wh10_plot}c and Fig.~\ref{fig:supp_wh16_plot}c. More generally, we expect the fidelity of an entangled state after a noisy channel can depend on the amount of entanglement in the input state, as noted in~\cite{Apollaro_2022}.

\begin{figure*}[h]
\centering
\includegraphics[width=\textwidth]{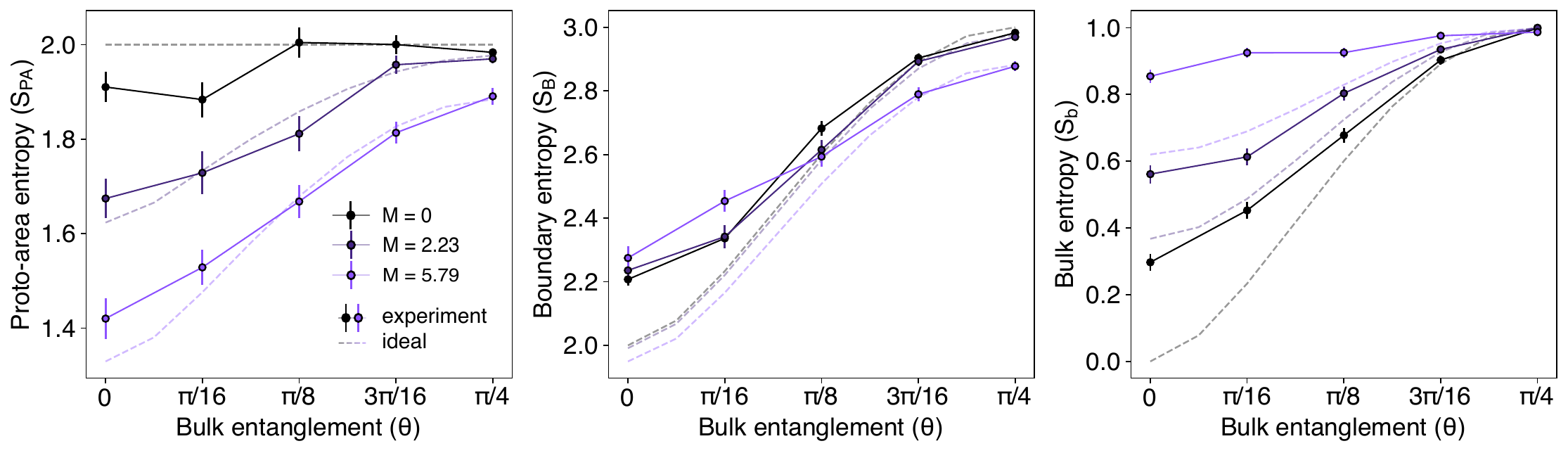}
\caption{
\textbf{Experiment data for the single-copy $[[8,2,3]]$ code using unequal depth circuit.}
The left panel shows the proto-area entropy as a function of the bulk entanglement angle $\theta$ for the same injected magic amounts as in the main text. The middle and right panels show the corresponding boundary entropy $S_{\mathrm{B}}$ and bulk entropy $S_{\mathrm{b}}$, respectively. Experimental points are obtained from QST with $1500$, $1250$, and $1000$ shots per Pauli basis for $M=0$, $M=2.23$, and $M=5.79$, respectively. Dashed lines show entropies from ideal, noiseless results.
}
\label{fig:supp_sc8_noisy_system_plot}
\end{figure*}

However, an important distinction from coherent noise is that the $\theta$-dependent entropy shifts caused by stochastic noise can largely cancel when the boundary and recovered bulk entropies are subtracted to form the proto-area entropy. This cancellation is most effective when the boundary-entropy and bulk-entropy measurements are performed using circuits with the same depth and noise structure. For example, in the single-copy $[[8,2,3]]$ code experiment, we measure the boundary entropy using the same circuit as the bulk entropy measurement by appending the recovery unitary before measurement. In this case, the boundary and bulk measurements experience the same stochastic noise, and their difference yields a nearly flat zero-magic proto-area curve, as shown in Fig.~\ref{fig:supp_sc8_plot}a. This indicates that the leading $\theta$-dependent contribution from stochastic noise is cancelled in $S_{\mathrm{PA}}$. 

By contrast, we perform another single-copy experiment to demonstrate how unequal circuit depths can generate an apparent $\theta$-dependent contribution to the proto-area entropy, as shown in Fig.~\ref{fig:supp_sc8_noisy_system_plot}. The boundary-entropy circuits do not include the recov-3 block and contain only $18$ native $ZZ$ gates, while the bulk-entropy circuits include it and contain $23$ native $ZZ$ gates. Because the two entropy measurements accumulate different amounts of noise, the cancellation in $S_{\mathrm{PA}}$ is incomplete. This produces a residual $\theta$-dependent slope even in the $M=0$ experiment, illustrating that such slopes can arise from noisy systematics rather than from injected magic alone.

\begin{figure}
    \centering
    \includegraphics[width=1.0\linewidth]{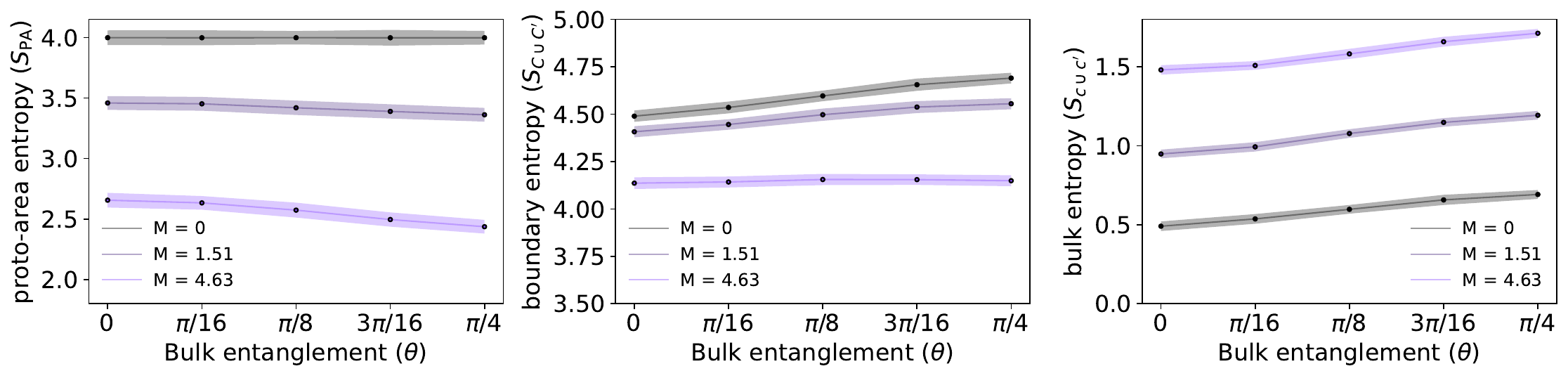}
    \caption{\textbf{Noisy numerical simulation for two-sided $[[5,1,3]]$ code using same circuit for the boundary and bulk entropy.}  }
    \label{fig:same_circuit_wh10}
\end{figure}

In the two-sided experiment, we did not use the same circuits for the boundary- and bulk-entropy measurements. Instead, we measured the boundary entropy from the complementary region, because this requires quantum state tomography only on a 4-qubit subregion, substantially reducing the experimental resource.  As a result, the stochastic noise introduced by the recovery circuit in the bulk-entropy measurement is not matched by a corresponding contribution in the boundary-entropy measurement, and the cancellation is incomplete. This explains why the zero-magic proto-area entropy is not perfectly flat in the two-sided experiment, as shown in Fig.~\ref{fig:supp_wh10_plot}a and Fig.~\ref{fig:supp_wh16_plot}a.

\begin{figure}
    \centering
    \includegraphics[width=1\linewidth]{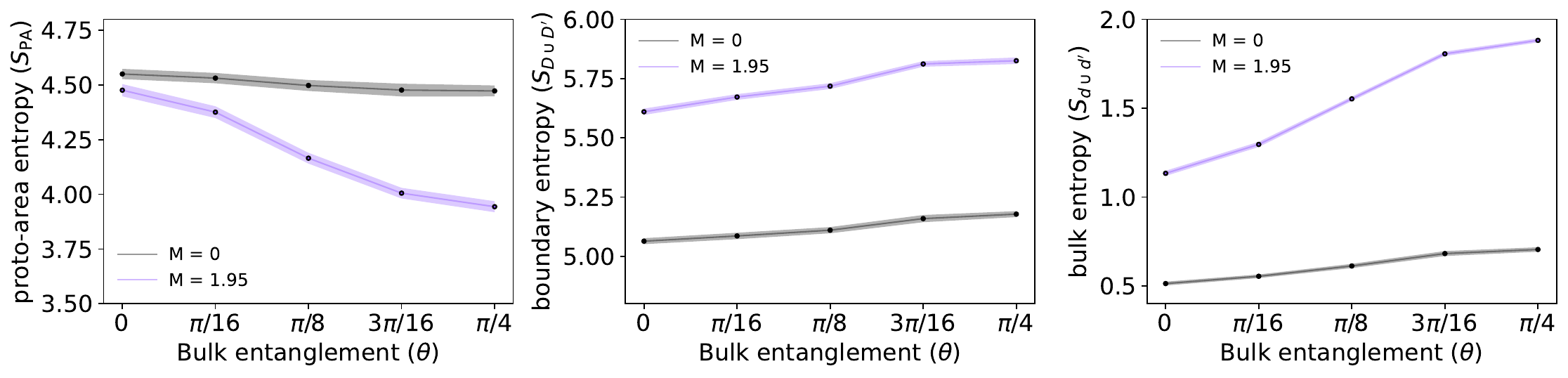}
    \caption{\textbf{Noisy numerical simulation for two-sided $[[8,2,3]]$ code using same circuit for the boundary and bulk entropy.} }
    \label{fig:same_circuit_wh16}
\end{figure}

To test this interpretation, we performed a noisy numerical simulation of the two-sided $[[5,1,3]]$ experiment using matched circuits for the boundary and bulk entropy estimation --- the boundary reduced density matrix is obtained from the 6-qubit subregion $C\cup C'$ after applying the recovery unitary. As shown in Fig.~\ref{fig:same_circuit_wh10}, the zero-magic proto-area curve becomes nearly flat. This supports the conclusion that the non-flat zero-magic curve in the original two-sided protocol arises primarily from unmatched stochastic noise contributions.  The same matched circuit simulation is also performed for two sided $[[8,2,3]]$ code and the entropy plots are shown in Fig.~\ref{fig:same_circuit_wh16}. 

We emphasize that this cancellation provides an important way to distinguish coherent and stochastic errors. Coherent gate-angle offsets act as fixed circuit perturbations and effectively inject additional magic, thereby changing the physical $\theta$ dependence of $S_{\mathrm{PA}}$. In contrast, stochastic noise mainly produces systematic entropy shifts that can be reduced, or largely cancelled, by using matched measurement circuits for the boundary and recovered bulk entropies.

\section{Numerical baseline with optimal recovery}\label{app:optimal}
\begin{figure}
    \centering
    \includegraphics[width=1.0\linewidth]{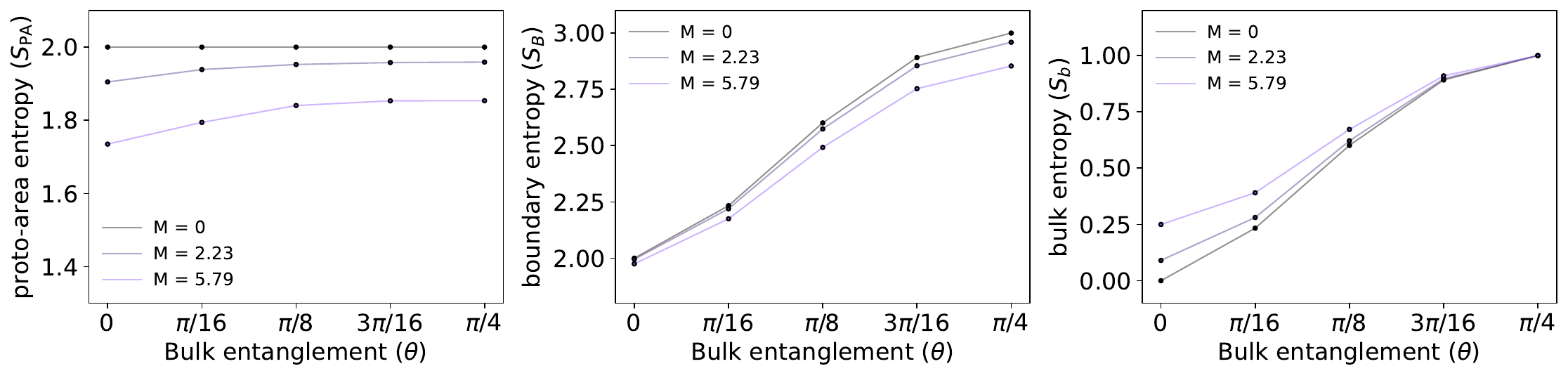}
    \caption{\textbf{Numerical simulation for the single-copy $[[8,2,3]]$ code using optimal recovery.} The left panel shows the proto-area entropy as a function of the bulk entanglement angle $\theta$. The middle and right panels show the corresponding boundary entropy $S_B$ and bulk entropy $S_b$, respectively. }
    \label{fig:optimal_823}
\end{figure}
In this section, we compute the proto-area entropy using the unconstrained optimal recovery as required by the theory definition in Ref.~\cite{cao2026} in order to establish a clean theoretical baseline for state-dependence that is unobscured by resource constraints and noise. As discussed in Appendix~\ref{app:recovery}, the unconstrained optimal recovery $R$ is obtained by minimizing the cost function $f(R)$ defined in Eq.~\eqref{eq:cost} over all unitaries in $SU(d_A)$, where $d_A$ is the Hilbert space dimension of the boundary subregion $A$, while the Clifford-architecture recovery is optimized only over the unitaries with the same circuit architecture as the  Clifford recovery. 

\begin{figure}[t]
    \centering

    \begin{subfigure}{0.45\textwidth}
        \centering
        \includegraphics[width=0.4\linewidth]{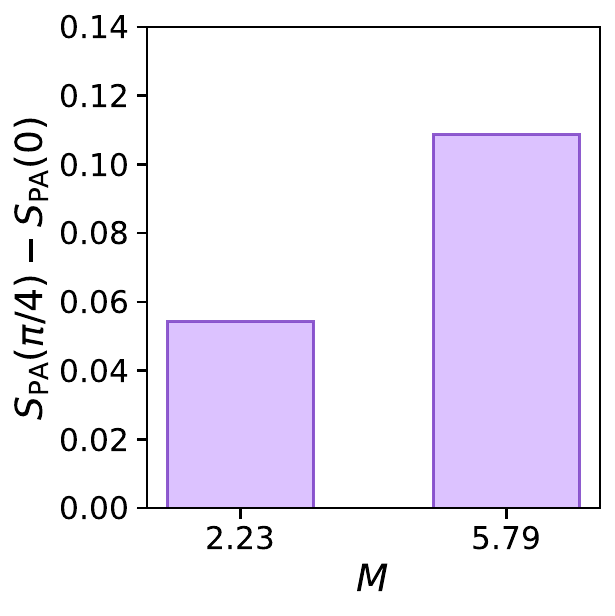}
        \caption{}
        \label{fig:barplota}
    \end{subfigure}
    \hspace{0.05\textwidth}
    \begin{subfigure}{0.45\textwidth}
        \centering
        \includegraphics[width=0.4\linewidth]{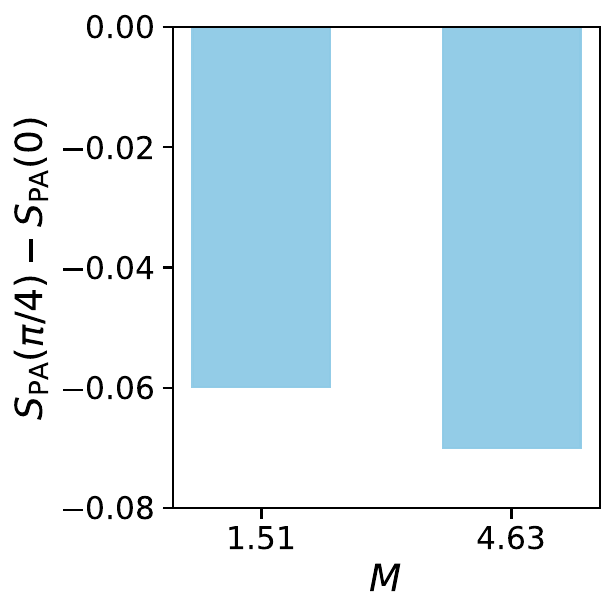}
        \caption{}
        \label{fig:barplotb}
    \end{subfigure}

    \caption{\textbf{Plot of proto-area entropy difference $S_{\mathrm{PA}}(\pi/4)-S_{\mathrm{PA}}(0)$.} (a) shows the proto-area entropy difference for the single-copy $[[8,2,3]]$ code and (b) shows the proto-area entropy difference for the two-sided $[[5,1,3]]$ code. This difference is exactly zero at $M=0$, the  stabilizer limit }
    \label{fig:main}
\end{figure}

Fig.~\ref{fig:optimal_823} shows the proto-area entropy plot for the single-copy $[[8,2,3]]$ code calculated numerically using the unconstrained optimal recovery. The positive correlation between the proto-area entropy and bulk entanglement angle $\theta$ still persists, although the proto-area difference $\Delta S_{\mathrm{PA}}:=S_{\mathrm{PA}}(\pi/4)-S_{\mathrm{PA}}(0)$ is reduced to $0.108$ for $M=5.79$ and $0.052$ for  $M=2.23$, respectively, as shown in Fig.~\ref{fig:barplota}. These $\Delta S_{\mathrm{PA}}$ are about four times smaller than those obtained using Clifford-architecture recovery.

\begin{figure}
    \centering
    \includegraphics[width=1.0\linewidth]{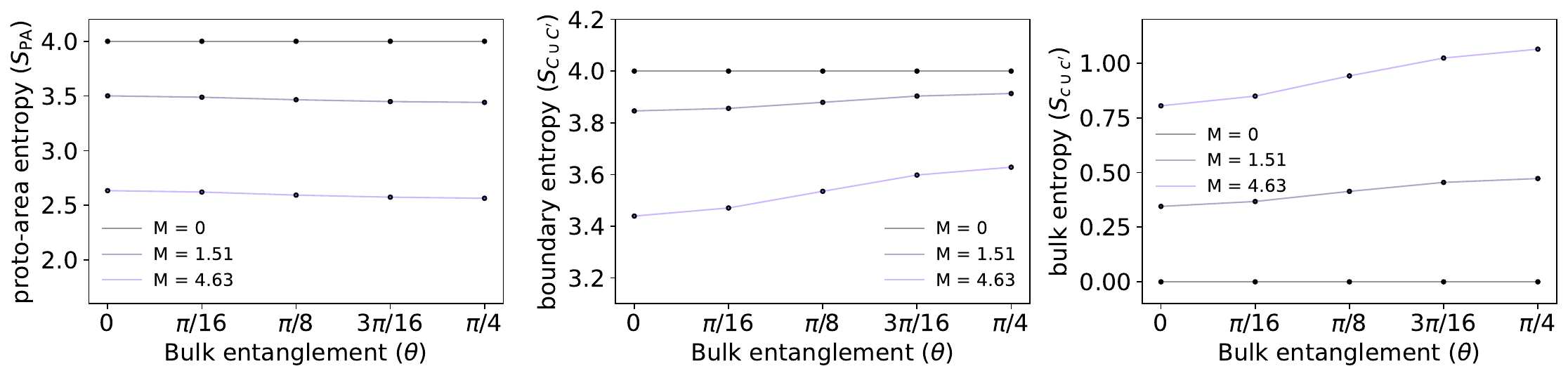}
    \caption{\textbf{Numerical simulation for the two-sided $[[5,1,3]]$ code using optimal recovery.} The left panel shows the proto-area entropy as a function of the bulk entanglement angle $\theta$. The middle and right panels show the corresponding boundary entropy $S_{C\cup C'}$ and bulk entropy $S_{c\cup c'}$, respectively.}
    \label{fig:optimal_wh10}
\end{figure}

Fig.~\ref{fig:optimal_wh10} shows the proto-area entropy plot for the two-sided $[[5,1,3]]$ code calculated numerically using the optimal recovery. The negative correlation between proto-area entropy and bulk entanglement again persists, with $\Delta S_{\mathrm{PA}}$ reduced to $-0.070$ for  $M=4.63$ and $-0.060$ for $M=1.51$, respectively, as shown in Fig.~\ref{fig:barplotb}. As in the single-copy case, these $\Delta S_{\mathrm{PA}}$ are also about four times smaller compared with those obtained from Clifford-architecture recovery.

This shows that the $\theta$ dependence is indeed tied to the choice of the recovery channel, and suboptimality amplifies the dependence in the entropy measurements. Nevertheless, the non-zero state dependence from the theoretical baseline is real and not merely a recovery-dependent artifact.

\section{Magic of the code}\label{app:NLmagic}
In this section, we quantify the amount of magic introduced to the perturbed code using stabilizer Rényi entropy (SRE)~\cite{Leone_2022} as a metric. Given an encoding circuit $V$, the amount of magic in the code~\cite{cao2026} can be defined to be the SRE of the Choi state of $V$. More specifically, denote the codeword state as $\ket{\tilde i}:=V\ket{i}_L$, logical Hilbert space dimension as $d_L$, and the SRE measure as $\mathcal{M}$. Then the magic of the code is 
\begin{equation}
    \mathcal{M}(V):= \mathcal{M}\left(\frac{1}{\sqrt{d_L}}\sum_{i=1}^{d_L}\ket{i}_R\ket{\tilde i}\right).
\end{equation}
Here $\ket{i}_R$ represents the basis state of the reference system used to define the Choi state. 
We next consider the non-local magic of the code. Given a partition $A$ and $A^c$, non-local magic involves minimizing the above SRE measure over local unitaries $U_A\otimes U_{A^c}$ acting on the Choi state. In practice, however, evaluating the SRE of a Choi state, with at least ten qubits, is already computationally costly; the full optimization is beyond the scope of the present work. Instead, for each codeword state $\ket{\tilde{i}}$, we calculate the non-local SRE \textit{estimate} defined in~\cite{Cao:2024nrx} and average the result over these codewords.

Tables~\ref{tab:magic_823}-\ref{tab:magic_16wh} summarize the computed magic values for different codes. In each table, the first column lists the over-rotation parameters for two-qubit $XX$-gate; the parameters for other gate-types are listed in Tables~\ref{tab:lambdas_823}-\ref{tab:lambdas_10wh}.

\begin{table}[H]
\centering
\caption{Magic of the single $[[8,2,3]]$ code for different values of $\lambda_{XX}$.}
\label{tab:magic_823}
\begin{tabular}{cccc}
\hline
Plot colors & $\lambda_{XX}$ & Total magic & Non-local magic \\
\hline
\colorsquare{000000} & $0$   & $0$      & $0$      \\
\colorsquare{A395BF} & $0.1$ & $2.2289$ & $0.0176$ \\
\colorsquare{C6A9FF} & $0.2$ & $5.7948$ & $0.0984$ \\
\hline
\end{tabular}
\end{table}

\begin{table}[H]
\centering
\caption{Magic of the two-sided $[[5,1,3]]$ code for different values of $\lambda_{XX}$.}
\label{tab:magic_10wh}
\begin{tabular}{cccc}
\hline
Plot colors & $\lambda_{XX}$ & Total magic & Non-local magic \\
\hline
\colorsquare{000000} & $0$     & $0$      & $0$      \\
\colorsquare{A395BF} & $0.135$ & $1.5148$ & $0.3218$ \\
\colorsquare{C6A9FF} & $0.27$  & $4.6326$ & $1.1608$ \\
\hline
\end{tabular}
\end{table}

\begin{table}[H]
\centering
\caption{Magic of the two-sided $[[8,2,3]]$ code for the optimized values of $\lambda_{i}$ within the range [-0.1, 0.1].}
\label{tab:magic_16wh}
\begin{tabular}{cccc}
\hline
Plot colors & $\lambda_{XX,\text{max}}$ & Total magic & Non-local magic \\
\hline
\colorsquare{000000} & $0$    & $0$      & $0$      \\
\colorsquare{C6A9FF} & $0.10$ & $1.9541$ & $0.1219$ \\
\hline
\end{tabular}
\end{table}

We find that both the total magic and non-local magic of the code increase with the over-rotation parameter. We also observe that the estimated non-local magic of the two-sided $[[5,1,3]]$ code is larger than that of the two-sided $[[8,2,3]]$ code, even though the latter exhibits a larger proto-area entropy signal. This apparent mismatch can be understood from the fact that the signal depends not only on the amount of magic, but also on the amount of bulk entanglement. In particular, the two-sided $[[8,2,3]]$ code encodes two Bell pairs, whereas the two-sided $[[5,1,3]]$ code encodes only one. We therefore expect the former to produce a larger proto-area entropy signal at a comparable level of magic. Additionally, $[[5,1,3]]$ and $[[8,2,3]]$ codes also have distinct structures, where a direct comparison is ill-defined.

More importantly, our estimate uses the average non-local magic of codeword states as a proxy for the non-local magic of the code. For a direct comparison between different codes, it would be preferable to use an intrinsic non-local magic measure of the encoding map itself. However, this ideal quantity is difficult to evaluate for the 10-qubit and 16-qubit circuits using current algorithms. We therefore leave the development of more computationally accessible measures of code non-local magic to future work.

\section{Entanglement and wormhole length}\label{app:Entanglement and wormhole length}

This section provides a heuristic holographic interpretation of the decrease in proto-area (PA) entropy observed in the two-sided code experiment. We argue that this trend is qualitatively consistent with holographic intuition: increasing the entanglement of quantum matter can enhance spacetime connectivity, as suggested by the ER=EPR proposal~\cite{Maldacena:2013xja,Jensen:2013ora}. We emphasize, however, that our model should be viewed as a holographic toy model at effectively $N=1$. It does not possess a semiclassical bulk dual, which is expected to emerge only in an appropriate large-$N$ limit.

In the two-sided model using the original HaPPY code, the emergent geometry is not affected by changes in the bulk entanglement. This can be understood by considering two disjoint boundary subregions, one on each side. The corresponding RT surface remains a disconnected union of the two individual RT surfaces, independent of the bulk state. This rigidity is consistent with the stabilizer nature of the HaPPY code, where the geometry is fixed by the tensor-network structure rather than dynamically affected by the encoded state.

In the single-sided setup, we observed that injecting magic into the HaPPY code induces state-dependent behavior in the emergent geometry. This motivates the conjecture that magic injection in the two-sided code can similarly make the effective geometry responsive to changes in bulk entanglement, thereby providing a circuit-level probe of the ER=EPR intuition.

Related theoretical works have explored similar ideas. Starting from two independent holographic CFTs with disconnected bulk duals, suitable inter-CFT couplings can lead to connected wormhole geometries~\cite{VanRaamsdonk:2009ar,VanRaamsdonk:2010pw,VanRaamsdonk:2018zws,Haehl:2019fjz,VanRaamsdonk:2020tlr,Sahu:2024ccg,Maloney:2025ina}. Related constructions based on random entanglement also show that entangled holographic systems can contain additional bulk regions in their entanglement wedge~\cite{Antonini:2023hdh}.

It is well-known that when the two sides of a holographic CFT are sufficiently entangled, the RT curve can switch from a pair of disconnected geodesics, each ending on one side, to a connected geodesic passing through the wormhole. Moreover, increasing the entanglement further shortens the connected geodesic. 
Following Ref.~\cite{Morrison:2012iz}, we revisit this observation in the context of our experiment.

Consider the thermofield-double (TFD) state
\begin{equation}
\ket{\mathrm{TFD}(\beta)}
=
\frac{1}{\sqrt{Z(\beta)}}
\sum_n
e^{-\beta E_n/2}
\ket{E_n}_L\ket{E_n}_R ,
\label{eq:tfd_state}
\end{equation}
where $\ket{E_n}_L$ and $\ket{E_n}_R$ denote energy eigenstates of the left and right CFTs, respectively, and $\beta=1/T$ is the inverse temperature. The normalization is fixed by the thermal partition function, 
\begin{equation}
Z(\beta)=\sum_n e^{-\beta E_n}.
\end{equation}
Tracing out either CFT leaves a thermal density matrix on the other side; the left--right entanglement entropy of the TFD state is therefore equal to the thermal entropy,
\begin{equation}
S_{\mathrm{TFD}}(\beta)
=
\left(1-\beta\frac{\partial}{\partial\beta}\right)
\log Z(\beta).
\label{eq:tfd_entropy}
\end{equation}
For a thermally stable state, this entropy increases with temperature according to
\begin{equation}
\frac{dS_{\mathrm{TFD}}}{dT}
=
\frac{C(T)}{T}>0 ,
\label{eq:entropy_heat_capacity}
\end{equation}
where $C(T)$ is the heat capacity. Thus, when the heat capacity is positive, decreasing $\beta=1/T$ increases the left--right entanglement entropy.

In the high-temperature Cardy regime of a two-dimensional holographic CFT, the thermal partition function takes the asymptotic form
\begin{equation}
\log Z(\beta)
\simeq
\frac{\pi c L_{\mathrm{CFT}}}{6\beta},
\label{eq:cardy_partition_function}
\end{equation}
where $c$ is the central charge and $L_{\mathrm{CFT}}$ is the spatial size of the CFT~\cite{Strominger:1997eq,Maldacena:2001kr}. Substituting Eq.~\eqref{eq:cardy_partition_function} into Eq.~\eqref{eq:tfd_entropy} gives
\begin{equation}
S_{\mathrm{TFD}}(\beta)
\simeq
\frac{\pi c L_{\mathrm{CFT}}}{3\beta}.
\label{eq:cardy_entropy}
\end{equation}
This entropy is equivalently the Bekenstein--Hawking entropy of the dual BTZ black hole. This relation shows explicitly that decreasing $\beta$ increases the left--right entanglement.

On the other hand, decreasing $\beta$ also shortens the geodesic passing through the BTZ wormhole. The regularized equal-time geodesic length connecting points $x_L$ and $x_R$ on the two opposite boundaries is~\cite{Morrison:2012iz}
\begin{equation}
L_{\mathrm{wh}}(x_L,x_R;\beta)
=
2\ell_{\mathrm{AdS}}
\log\left[
\frac{\beta}{\pi\epsilon}
\cosh\left(
\frac{\pi |x_L-x_R|}{\beta}
\right)
\right],
\label{eq:btz_cross_boundary_length_beta}
\end{equation}
where $\epsilon$ is the UV cutoff, $\ell_{\mathrm{AdS}}$ is the AdS radius, and $x_L$ and $x_R$ denote the boundary coordinates on the left and right CFTs, respectively. For aligned insertions, $x_L=x_R$, this expression reduces to
\begin{equation}
L_{\mathrm{wh}}(\beta)
=
2\ell_{\mathrm{AdS}}
\log\left(
\frac{\beta}{\pi\epsilon}
\right).
\label{eq:btz_aligned_length}
\end{equation}
Thus lowering \(\beta\), or equivalently increasing the left--right entanglement, shortens the regularized geodesic length. Combining Eqs.~\eqref{eq:cardy_entropy} and~\eqref{eq:btz_aligned_length}, the aligned geodesic length can be expressed in terms of the left--right entanglement entropy as
\begin{equation}
L_{\mathrm{wh}}(S_{\mathrm{TFD}})
=
2\ell_{\mathrm{AdS}}
\log\left(
\frac{cL_{\mathrm{CFT}}}{3\epsilon\,S_{\mathrm{TFD}}}
\right),
\label{eq:wormhole_length_entropy}
\end{equation}
up to the same cutoff-dependent additive constant $\epsilon$. This makes explicit the negative correlation   between the left--right entanglement and the regularized wormhole length in the holographic CFT calculation.   

In the $\mathrm{AdS}_3/\mathrm{CFT}_2$ correspondence, spacelike geodesics play the role of extremal surfaces anchored on the boundary. Consider a boundary region consisting of two intervals, one on each side of the TFD state, $A_L\cup A_R\equiv [x_L,y_L]\cup [x_R,y_R]$. There are two competing RT configurations. The first is a disconnected configuration, consisting of one geodesic joining $x_L$ to $y_L$ on the left boundary and another joining $x_R$ to $y_R$ on the right boundary. The second is a connected configuration, consisting of two geodesics that pass through the wormhole and connect the endpoints across the two boundaries, pairing $x_L$ with $x_R$ and $y_L$ with $y_R$.

Denote the size of the boundary interval as $\ell_{L/R}\equiv |y_{L/R}-x_{L/R}|$. When ${(\ell_L+\ell_R)}/{\beta}\gg 1$, the disconnected geodesic has minimal length and serves as the RT surface, while for small ${(\ell_L+\ell_R)}/{\beta}\ll 1$ the connected geodesic has minimal length and becomes the RT surface for the combined boundary intervals. This transition between two sets of geodesics can be captured by the holographic thermo-mutual information (HTMI)~\cite{Morrison:2012iz},
\begin{align}
\mathrm{HTMI}(A_L:A_R)
&=
\frac{1}{4G_N}
\max\Big\{
|\mathcal{L}_{L}|
+
|\mathcal{L}_{R}|
-
|\mathcal{L}_C|,
0
\Big\}.
\label{eq:htmi_rt}
\end{align}
Here \(|\mathcal{L}|\) denotes the length of the geodesic \(\mathcal{L}\). The geodesics $\mathcal{L}_{L}$ and $\mathcal{L}_{R}$ are the disconnected one-sided RT surfaces homologous to $A_L$ and $A_R$, respectively, while \(\mathcal{L}_C\) denotes the connected pair of geodesics passing through the wormhole and terminating at the end points of \(A_L\cup A_R\). The connected configuration dominates when 

\begin{equation}
|\mathcal{L}_C|
<
|\mathcal{L}_{L}|
+
|\mathcal{L}_{R}| .
\label{eq:connected_surface_condition}
\end{equation}
\begin{figure}[H]
    \centering
    \includegraphics[width=0.5\linewidth]{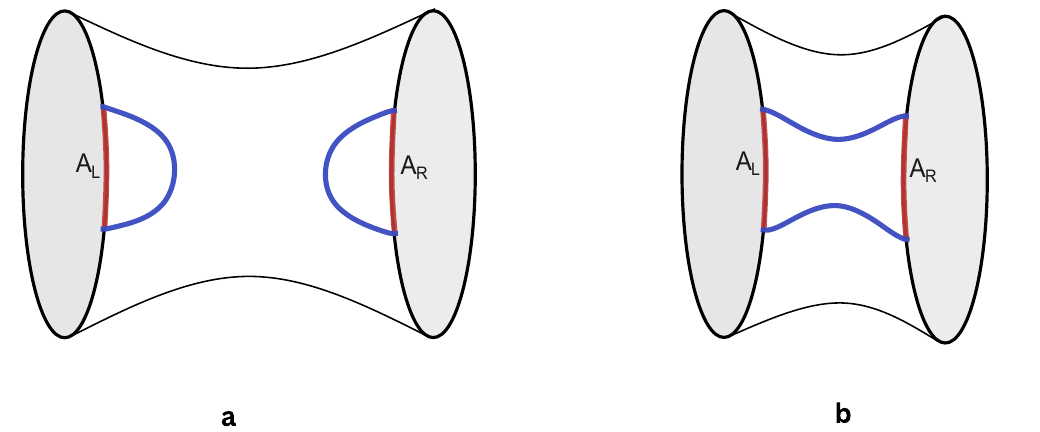}
    \caption{\textbf{Disconnected and connected RT configurations in a two-sided geometry.} For a boundary region $A_L\cup A_R$ consisting of one interval on each boundary, there are two competing extremal-surface configurations. Red segments denote the boundary intervals and blue curves denote candidate RT surfaces. (a) At low left--right entanglement, the minimal surface is the disconnected union $\mathcal L_L\cup \mathcal L_R$, with one geodesic homologous to each boundary interval. (b) Above the critical two-sided entanglement, the connected configuration $\mathcal L_C$ passing through the wormhole has smaller total length and computes the entropy of $A_L\cup A_R$. Further increasing the entanglement, equivalently decreasing $\beta$ in the TFD description, shortens the connected cross-wormhole geodesics.}
\label{fig:connected_geometry}
\end{figure}
Thus, once the two-sided entanglement is sufficiently large, the entropy of the combined boundary region is computed by a connected RT surface passing through the wormhole, rather than by two disconnected RT surfaces on the two sides.

For equal size boundary intervals, $\ell_L=\ell_R\equiv \ell$,  on the two boundaries, with aligned boundary coordinate $x_L=x_R$, the HTMI simplifies to~\cite{Morrison:2012iz}
\begin{equation}
\mathrm{HTMI}(A_L:A_R)
=
\frac{c}{12}
\log \left(
\max\left[
\frac{
\left(1-\cosh\frac{\pi\ell}{\beta}\right)^2
}{4},
1
\right]\right).
\label{eq:htmi_equal_intervals}
\end{equation}
The transition in RT surfaces occurs at
\begin{equation}
\beta_c
=
\frac{\pi\ell}{\cosh^{-1}3}.
\label{eq:beta_critical_htmi}
\end{equation}

Substituting in Eq.~\eqref{eq:cardy_entropy} gives a critical value of two-sided entanglement 
\begin{equation}
    S_c= c\frac{\cosh^{-1}{3} }{3}\frac{L_{\mathrm{CFT}}}{\ell}.
\end{equation}

Above this critical entanglement, the connected geodesic becomes the RT surface. Further increasing the two-sided entanglement, equivalently decreasing $\beta$, then shortens the connected geodesics passing through the wormhole.

Remarkably, the critical value is  independent of the UV cutoff. It depends instead on the central charge $c$, which measures the number $N$ of local degrees of freedom in the large-$N$ CFT. This suggests that, in principle, the transition can be meaningfully probed in a suitably regularized large-$N$ model.

Finally, we emphasize that the two-sided code construction studied in the experiment should be viewed as an $N=1$ toy model. While the decrease of proto-area (PA) entropy with increasing inter-code (bulk) entanglement is similar to the heuristic holographic interpretation on a high level, we comment on two important distinctions from holographic theories with semiclassical bulk duals. 

First is the absence of a geodesic-switching phase transition. In our finite-size circuit model, the PA entropy varies smoothly with bulk entanglement; however, we conjecture that in larger-$N$ generalizations this crossover may sharpen into a genuine phase transition, characterized by a non-analytic change in the derivative of the PA entropy. It would be interesting to investigate this possibility experimentally using multiple copies of the two-sided code, as well as analytically through tensor-network and Weingarten techniques along the lines of Ref.~\cite{cao2026}.

Second, a smooth geometry in holography is unlikely to emerge in our experiment, or indeed, in even larger systems with arbitrarily applied entanglement patterns. It is well-known that complex entanglement patterns have to satisfy stringent constraints on their entropies before a consistent geometric interpretation can arise, see e.g., \cite{Bao_2015,Marolf_2017,Bao_2019,Cao_2020}.

\end{document}